\documentclass[iop,revtex4]{emulateapj}

\shortauthors{LAPI ET AL.}
\shorttitle{STELLAR MASS FUNCTION VIA CONTINUITY EQUATION}

\begin{document}

\title{Stellar Mass Function of Active and Quiescent Galaxies \\
via the Continuity Equation}
\author{A. Lapi\altaffilmark{1,2,3}, C. Mancuso\altaffilmark{4,2}, A. Bressan\altaffilmark{1,2,3}, L. Danese\altaffilmark{1,2,3}}
\altaffiltext{1}{SISSA, Via Bonomea 265, 34136 Trieste, Italy}\altaffiltext{2}{INFN-Sezione di Trieste, via Valerio 2, 34127 Trieste, Italy}\altaffiltext{3}{INAF-Osservatorio Astronomico di Trieste, via Tiepolo 11, 34131 Trieste, Italy}
\altaffiltext{4}{INAF-IRA, Via P. Gobetti 101, 40129 Bologna, Italy}
\begin{abstract}
The continuity equation is developed for the stellar mass content of galaxies, and exploited to derive the stellar mass function of active and quiescent galaxies over the redshift range $z\sim 0-8$. The continuity equation requires two specific inputs gauged on observations: (i) the star formation rate functions determined on the basis of the latest UV+far-IR/sub-mm/radio measurements; (ii) average star-formation histories for individual galaxies, with different prescriptions for discs and spheroids. The continuity equation also includes a source term taking into account (dry) mergers, based on recent numerical simulations and consistent with observations. The stellar mass function derived from the continuity equation is coupled with the halo mass function and with the SFR functions to derive the star formation efficiency and the main sequence of star-forming galaxies via the abundance matching technique. A remarkable agreement of the resulting stellar mass function for active and quiescent galaxies, of the galaxy main sequence and of the star-formation efficiency with current observations is found; the comparison with data also allows to robustly constrain the characteristic timescales for star formation and quiescence of massive galaxies, the star formation history of their progenitors, and the amount of stellar mass added by in-situ star formation vs. that contributed by external merger events. The continuity equation is shown to yield quantitative outcomes that must be complied by detailed physical models, that can provide a basis to improve the (sub-grid) physical recipes implemented in theoretical approaches and numerical simulations, and that can offer a benchmark for forecasts on future observations with multi-band coverage, as it will become routinely achievable in the era of JWST.
\end{abstract}

\keywords{galaxies: evolution --- galaxies: statistics --- galaxies: star formation  --- galaxies: luminosity function, mass function --- galaxies: high redshift}

\setcounter{footnote}{0}

\section{Introduction}\label{sec|intro}

Some recent findings have significantly rekindled interest in the field of galaxy formation and evolution. The first concerns the discovery of an abundant population of dusty star-forming galaxies at redshifts $z\ga 1$, that has been shown to be responsible for the bulk of the cosmic star formation history, in particular around the crucial redshifts $z\approx 2-3$ where it peaks (e.g., Gruppioni et al. 2013; Rowan-Robinson et al. 2016; Lapi et al. 2017; Bourne et al. 2017; Dunlop et al. 2017; Novak et al. 2017), and to be present even out to $z\la 6$ (e.g., Cooray et al. 2014; Riechers et al. 2017; Zavala et al. 2017). Such achievement has become feasible only recently thanks to wide-area far-IR/sub-mm surveys conducted by \textsl{Herschel}, ASTE/AzTEC, APEX/LABOCA, JCMT/SCUBA2, and ALMA-SPT (e.g., Gruppioni et al. 2013, 2015; Lapi et al. 2011; Weiss et al. 2013; Strandet et al. 2016; Koprowski et al. 2014, 2016), in many instances eased by gravitational lensing from foreground objects (e.g., Negrello et al. 2014, 2017; Nayyeri et al. 2016). In fact, galaxies endowed with star formation rates $\dot M_\star\ga$ a few tens $M_\odot$ yr$^{-1}$ at redshift $z\ga 2$ were largely missed by rest-frame optical/UV surveys because of heavy dust obscuration, difficult to correct for with standard techniques based only on UV spectral data (e.g., Bouwens et al. 2016, 2017; Mancuso et al. 2016a; Pope et al. 2017; Ikarashi et al. 2017; Simpson et al. 2017).

High-resolution, follow-up observations of these galaxies in the far-IR/sub-mm/radio band via ground-based interferometers, such as SMA, VLA, PdBI, and recently ALMA, have revealed star formation to occur in a few collapsing clumps distributed over spatial scales smaller than a few kpcs (see Simpson et al. 2015; Ikarashi et al. 2015; Straatman et al. 2015; Spilker et al. 2016; Barro et al. 2016; Tadaki et al. 2017). A strongly baryon-dominated stellar core with high ongoing SFR is often surrounded out to $\la 15$ kpc by a clumpy, unstable gaseous disk in nearly keplerian rotation (e.g., Genzel et al. 2017; Swinbank et al. 2017).

Observations of dusty star-forming galaxies in the optical and near/mid-IR band from \textsl{Spitzer}, \textsl{WISE}, and \textsl{HST} have allowed to characterize their stellar mass content. The vast majority feature stellar masses strongly correlated to the SFR, in the way of an almost linear relationship dubbed 'Main Sequence', with a normalization
steadily increasing as a function of redshift and a limited scatter around $0.25$ dex (see Daddi et al. 2007; Rodighiero et al. 2011, 2015; Speagle et al. 2014; Whitaker et al. 2014; Renzini \& Peng 2015; Salmon et al. 2015; Tasca et al. 2015; Kurczynski et al. 2016; Tomczak et al. 2016; Bourne et al. 2017; Dunlop et al. 2017; Schreiber et al. 2017).

Another relevant piece of news concerns the discovery by deep near-IR surveys of an increasing number of massive galaxies $M_\star\ga$ several $10^{10}\, M_\odot$ at high redshift $z\ga 2$ (see Bernardi et al. 2013, 2017; Ilbert et al. 2013; Duncan et al. 2014; Tomczak et al. 2014; Caputi et al. 2015; Grazian et al. 2015; Thanjavur et al. 2016; Song et al. 2016; Davidzon et al. 2017). Even more interestingly, some of them are found to be already in passive evolution at $z\ga 2-3$, and to feature chemical properties similar to local early-type galaxies, including a (super)solar metallicity and a pronounced $\alpha$-enhancement. There is the intriguing yet still debated possibility that the dusty star-forming objects seen in the far-IR/sub-mm band constitute the progenitors of the massive (quiescent) galaxies increasingly detected at high redshifts via deep near-IR surveys (Straatman et al. 2014, 2016; Lonoce et al. 2015; Kriek et al. 2016; Mawatari et al. 2016; Michalowski et al. 2016; Davidzon et al. 2017; Glazebrook et al. 2017).

Relevant model-independent information on the cosmic star formation and mass growth history can be inferred by comparing the observed SFR function, stellar mass function, and main sequence for active and quiescent galaxies (e.g., Leja et al. 2015; Contini et al. 2016; Tomczak et al. 2016; Mancuso et al. 2016a,b; Steinhardt et al. 2017). This procedure can provide stringent constraints, e.g., on the typical timescales for star formation and quiescence, on the overall star formation efficiency, on the initial mass function (IMF), and on the amount of stellar mass added by in-situ star formation vs. that contributed by external merger events. Such outcomes can also be helpful to improve the (sub-grid) physical recipes implemented in theoretical models and numerical simulations, that currently face some difficulties in reproducing the observed abundances of strongly star-forming and massive quiescent galaxies at $z\ga 2-3$ (e.g., Wellons et al. 2015; Behroozi \& Silk 2017; Dave et al. 2017; Furlong et al. 2017; Rong et al. 2017; Hopkins et al. 2017).

In the present paper we pursue the above strategy, for the first time in a \emph{quantitative} way, by exploiting the specific tool constituted by the `continuity equation'. Being originally devised to connect quasar statistics to the demographics of supermassive black hole relics (Cavaliere et al. 1971; Soltan 1982; Small \& Blandford 1992; Salucci et al. 1999; Yu \& Lu 2004, 2008; Marconi et al. 2004; Merloni \& Heinz 2008; Shankar et al. 2009, 2013; Aversa et al. 2015), here we develop it for the stellar mass content of galaxies, in order to derive the stellar mass function of active and passive galaxies at different redshifts from the SFR functions and average star formation histories for individual objects. Our approach includes in the continuity equation a source term taking into account dry mergers and tidal stripping effects, gauged on observations and on state-of-the-art numerical simulations. With the term dry mergers we refer to events adding the whole mass content in stars of merging objects without contributing significantly to in-situ star formation; starbursts triggered by wet mergers, although included as star-forming objects populating the SFR functions, are expected to contribute little to the final stellar mass, and especially so for massive galaxies.

Moreover, we will exploit the abundance matching technique to derive the star formation efficiency and the main sequence of star-forming galaxies, and compare the outcome to recent observational determinations. Specifically, we will demonstrate via the continuity equation that the dusty, strongly star-forming galaxies at $z\ga 2$ are indeed the progenitors of massive quiescent galaxies, and that the latter's mass growth is dominated by in-situ star formation with an overall efficiency of less than $20\%$.

The plan of the paper is straightforward. In \S~\ref{sec|basics} we describe the basic ingredients of our analysis: the SFR functions and the adopted star formation histories for individual galaxies; in \S~\ref{sec|continuity} we solve the continuity equation for the stellar mass function of active and passive galaxies, and describe how to derive from those the star formation efficiency and the main sequence of star-forming galaxies; in \S~\ref{sec|results} we present our results and compare them to observations, discussing the relevant implications for galaxy formation and evolution; in \S~\ref{sec|summary} we summarize our findings.

Throughout the work we adopt the standard flat cosmology (Planck Collaboration XIII 2016) with round parameter values: matter density $\Omega_M = 0.32$, baryon density $\Omega_b = 0.05$, Hubble constant $H_0 = 100\, h$ km s$^{-1}$ Mpc$^{-1}$ with $h = 0.67$, and mass variance $\sigma_8 = 0.83$ on a scale of $8\, h^{-1}$ Mpc. Stellar masses and SFRs (or luminosities) of galaxies are evaluated assuming the Chabrier's (2003) IMF.

\section{Basic ingredients}\label{sec|basics}

Our analysis relies on two basic ingredients: (i) an observational determination of the SFR function at different redshifts; (ii) deterministic evolutionary tracks describing the average star formation history of individual galaxies. In this section we recall the notions relevant for the investigation of the stellar mass function, deferring the reader to the papers by Mancuso et al. (2016a,b) and Lapi et al. (2017) for more details.

\subsection{SFR functions and cosmic SFR density}\label{sec|SFR_func}

The first ingredient is constituted by the intrinsic SFR function ${\rm d}N/{\rm d}\log \dot M_\star$, namely the number density of galaxies per logarithmic bin of SFR $[\log \dot M_\star,\log\dot M_\star+{\rm d}\log\dot M_\star]$ at given redshift $z$. This has been accurately determined by Mancuso et al. (2016a,b) and Lapi et al. (2017) by exploiting the most recent determinations of the evolving galaxy luminosity functions from (dust-corrected) UV, far-IR, sub-mm, and radio data.

\begin{figure*}
\epsscale{1}\plotone{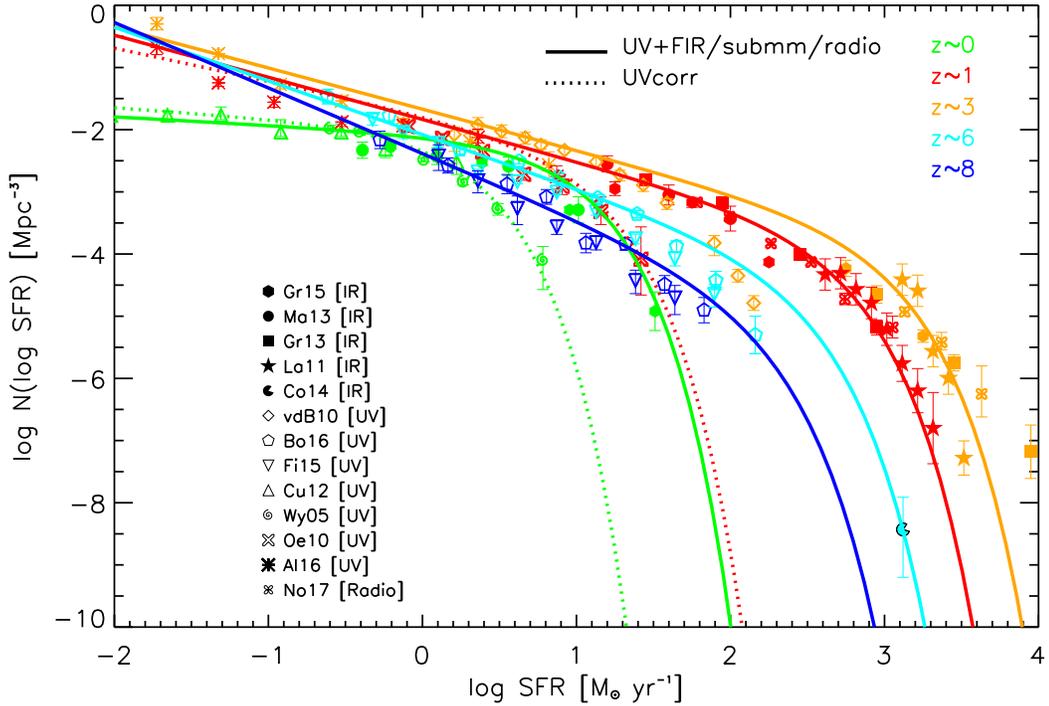}\caption{The SFR functions at redshifts $z=0$ (green), $1$ (red), $3$ (orange) and $6$ (blue) determined according to the procedure by Mancuso et al. (2016a,b) and Lapi et al. (2017). Solid lines refer to the rendition from UV plus far-IR/sub-mm/radio data; dotted lines (only plotted at $z\approx 0$ and $1$) refer to the rendition from UV data (dust corrected according to standard prescriptions based on the UV slope). UV data (open symbols) are from van der Burg et al. (2010; diamonds), Bouwens et al. (2016, 2017; pentagons), Finkelstein et al. (2015; inverse triangles), Cucciati et al. (2012; triangles), Wyder et al. (2005; spirals), Oesch et al. (2010; crosses), Alavi et al. (2016; asterisks); far-IR/sub-mm data from Gruppioni et al. (2015; hexagons), Magnelli et al. (2013; circles), Gruppioni et al. (2013; squares), Lapi et al. (2011; stars), and Cooray et al. (2014; pacmans); radio data from Novak et al. (2017; clovers).}\label{fig|SFR_func}
\end{figure*}

The SFR function can be described as a smooth Schechter function
\begin{equation}
{{\rm d}N\over {\rm d}\log\dot M_\star}(\dot M_\star,z) = \mathcal{N}(z)\, \left[\dot M_\star\over\dot M_{\star, c}(z)\right]^{1-\alpha(z)}\,e^{-\dot M_\star/\dot M_{\star, c}(z)}~,
\end{equation}
with three parameters: the overall normalization $\mathcal{N}$, the characteristic SFR $\dot M_{\star, c}$ and the faint end slope $\alpha$. The redshift evolution of each parameter has been measured via an educated fit to the observed data in unitary redshift bins by Mancuso et al. (2016a,b).
As extensively discussed by the latter authors, the SFR function is mainly determined by (dust-corrected) UV data for SFR $\dot M_\star\la 30\, M_\odot$ yr$^{-1}$ since in this range dust emission is mainly due to the diffuse (cirrus) dust component and standard UV dust-corrections based on the UV slope are reliable (see Meurer et al. 1999; Calzetti 2000; Bouwens et al. 2015, 2016, 2017); here we use the Meurer/Calzetti extinction law, but note that switching to a Small Magellanic Cloud (SMC) extinction law affects mildly the SFR function at the faint end (see also Sect.~\ref{sec|results} and Fig.~\ref{fig|MF_complot}). On the other hand, the SFR function is mainly determined by far-IR/sub-mm/radio data for SFRs $\dot M_\star\ga 10^2\, M_{\odot}$ yr$^{-1}$ since in this range dust emission is largely dominated by molecular clouds, and UV corrections are wildly dispersed and statistically fail (see Silva et al. 1998; Efstathiou et al. 2000; Coppin et al. 2015; Reddy et al. 2015; Fudamoto et al. 2017).

The resulting SFR functions at representative redshifts are illustrated along with the relevant data collection in Fig.~\ref{fig|SFR_func}. In Mancuso et al. (2016a,b; 2017) and Lapi et al. (2017) we have validated them against independent datasets, including integrated galaxy number counts at significative far-IR/sub-mm/radio wavelengths, counts/redshift distributions of strongly gravitationally-lensed galaxies, main sequence of star-forming galaxies and AGNs, redshift evolution of the cosmic SFR, and high-redshift observables including the history of cosmic reionization.

All in all, our determination of the SFR functions implies a significant number density of dusty star-forming galaxies with SFR $\dot M_\star\ga 10^2\, M_\odot$ yr$^{-1}$, currently missed by (dust-corrected) UV data. To highlight more clearly this point, in Fig.~\ref{fig|SFR_func} we also report at $z\la 1$ the SFR function that would have been inferred basing solely on UV data, dust corrected via the UV slope. The UV data considerably underestimate the SFR function for SFRs $\dot M_\star\ga 30\, M_\odot$ yr$^{-1}$, because of strong dust extinction. Interestingly, the shape of the SFR function for $\dot M_\star\ga 10^2\, M_\odot$ yr$^{-1}$, which so far has been probed only indirectly at $z\ga 4$ due to sensitivity limits in current wide-areas far-IR surveys, is found to agree out to $z\la 6$ with the constraints from the recent VLA-COSMOS radio survey (Novak et al. 2017) and from the few individual galaxies detected at $z\ga 5$ with ALMA and SMA (e.g., Riechers et al. 2017; Zavala et al. 2017). We shall demonstrate via the continuity equation that a robust probe on the bright end of the SFR function at high-redshift $z\ga 4$ is provided by the galaxy stellar mass function.

\begin{figure*}
\epsscale{1}\plotone{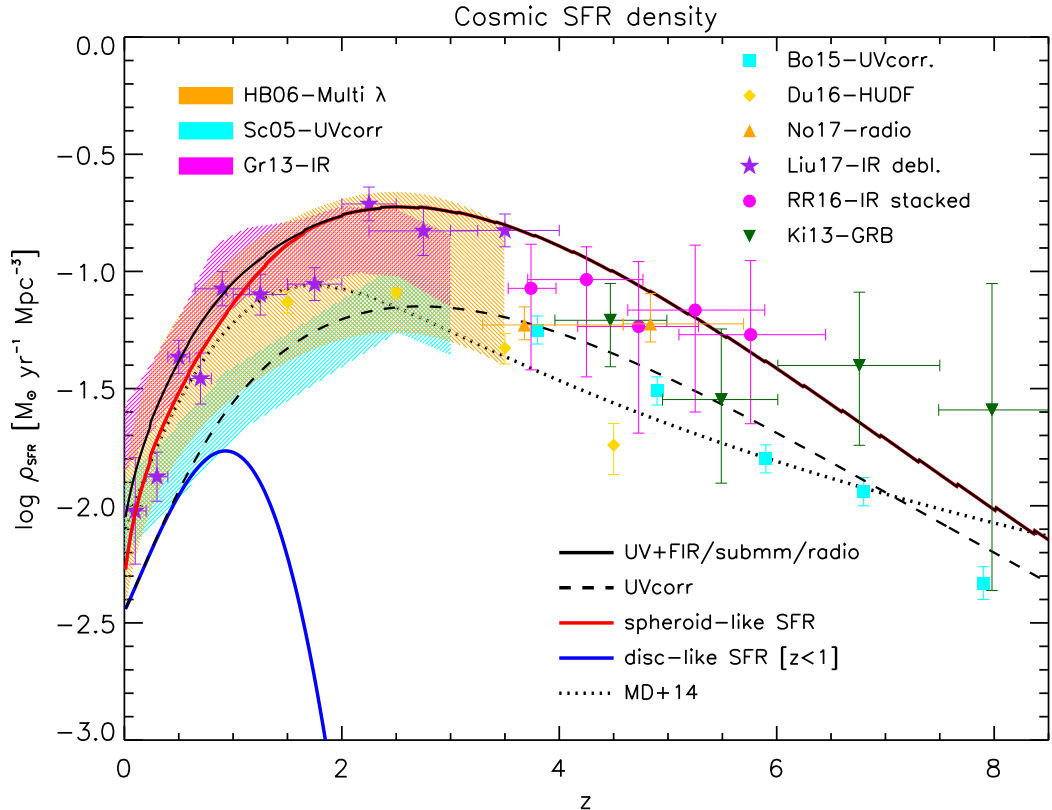}\caption{Cosmic SFR density as a function of redshift. The black solid line is the density derived from integrating the total (UV+far-IR/sub-mm/radio) SFR functions of Fig.~\ref{fig|SFR_func} down to a limit $\dot M_\star\sim 10^{-1}\, M_\odot$ yr$^{-1}$. The contribution from objects with spheroid-like and disc-like star-formation history (see Sect.~\ref{sec|basics}) are highlighted by the red and blue solid lines, respectively. The black dashed line illustrates the SFR density when basing on the (dust-corrected) UV-inferred SFR functions at any redshift. For reference, the dotted line illustrates the determination by Madau \& Dickinson (2014). Data are from: (dust-corrected) UV observations by Schiminovich et al. (2005; cyan shaded area) and Bouwens et al. (2015; cyan squares); ALMA sub-mm observations of UV-selected galaxies on the HUDF by Dunlop et al. (2016); VLA radio observations on the COSMOS field by Novak et al. (2017); multiwavelength determination including UV, radio, H$\alpha$, and mid-IR 24 $\mu$m data collected by Hopkins \& Beacom (2006; orange shaded area); \textsl{Herschel} far-IR observations by Gruppioni et al. (2013; magenta shaded area); \textsl{Herschel} far-IR stacking by Rowan-Robinson (2016; magenta circles); far-IR/sub-mm observations from deblended data on the GOODS field by Liu et al. (2017); estimates from long GRB rates by Kistler et al. (2009, 2013; green stars).}\label{fig|SFR_cosm}
\end{figure*}

For the analysis in the present paper, we shall assume that at $z\ga 1$ active galaxies populate the total (UV+far-IR/sub-mm/radio) SFR function, and feature a spheroid-like star-formation history; the latter envisages a nearly constant beahvior of the SFR as a function of galaxy age, with a timescale of $\la 1$ Gyr at high SFR $\dot M_\star\ga 30\, M_\odot$ yr$^{-1}$, increasing to a few Gyrs for lower SFRs (see Sect.~\ref{sec|SFR_hist} for details). At high SFRs such a population comprises dusty starforming, far-IR/sub-mm selected galaxies, which will turn out to be the progenitors of local massive dead spheroids with masses $M_\star\ga$ a few $10^{10}\, M_\odot$; at low SFRs, it comprises mildly obscured, UV selected galaxies (e.g., Lyman break galaxies), that will end up in objects with stellar masses $M_\star\la 10^{10}\, M_\odot$.

At $z\la 1$ we will show that a bimodal star-formation history is required. On the one hand, the bright end of the total (UV+far-IR/sub-mm/radio) SFR function is assumed to be populated by galaxies with the same spheroid-like star-formation history sketched above; since the knee of the SFR function toward $z\sim 0$ recedes a lot, the star-formation timescales are on average appreciably longer than at $z\sim 1$, attaining up to a few Gyrs (possibly splitted in many recurrent, shorter bursts); this population comprises a mixed bag of objects, including low-mass spheroids (e.g., bulges), irregulars, and reactivations of massive galaxies. On the other hand, the UV-inferred SFR function is assumed to be populated by galaxies with a disc-like star-formation history, i.e., exponentially declining SFR as a function of galaxy age with long timescales of several Gyrs (see also Cai et al. 2013, 2014); these objects will end up in disc-dominated galaxies with stellar masses $M_\star\la$ several $10^{10}\, M_\odot$. In Sect.~\ref{sec|SFR_hist} we will describe in detail the adopted spheroid-like or disc-like star-formation histories for individual galaxies.

From the SFR function, we can straightforwardly compute the cosmic SFR density as
\begin{equation}\label{eq|SFR_func}
\rho_{\rm SFR}(z) = \int{\rm d}\log \dot M_{\star}\, {{\rm d}N\over {\rm d}\log \dot M_\star}\, \dot M_\star~,
\end{equation}
integrated down to a limit $\dot M_\star\la 10^{-1}\, M_\odot$ yr$^{-1}$ for fair comparison to observational data, in particular with current blank-field UV surveys at high $z\ga 4$; the outcome is illustrated in Fig.~\ref{fig|SFR_cosm}. The result from the (dust-corrected) UV-inferred SFR functions is in good agreement with the UV data by Schiminovich et al. (2005) at $z\lesssim 4$ and by Bouwens et al. (2015, 2016, 2017) at $z\gtrsim 4$. It also agrees with the estimate by ALMA observations of UV-selected galaxies in the HUDF (see Dunlop et al. 2017); this is because the rather small area of the HUDF survey allows to pick up only moderately star forming galaxies with mild dust obscuration, on which the UV slope-based corrections still work pretty well.

However, the cosmic SFR density from (dust-corrected) UV data is inconsistent with other datasets both at low and high redshift. Specifically, at redshifts $z\la 4$ it falls short with respect to the multiwavelength determination by Hopkins \& Beacom (2006) based on UV/optical, radio, H$\alpha$ and mid-IR $24\, \mu$m data, to the far-IR measurements from \textsl{Herschel} by Magnelli et al. (2013) and Gruppioni et al. (2013), and to the recent estimate from deblended data from \textsl{Herschel}, JCMT/AzTEC and JCMT/SCUBA-2 in the GOODS field by Liu et al. (2017). At redshifts $z\ga 4$ it underestimates the determinations based on stacking of far-IR data from \textsl{Herschel} by Rowan-Robinson et al. (2016), the measurements from radio data by Novak et al. (2017), and the estimates based on long GRB rates from \textsl{Swift} by Kistler et al. (2009; 2013). This mostly reflects the fact, already mentioned above, that the UV-inferred SFR functions (even corrected for dust extinction via the UV slope) appreciably underestimate the number density of dusty galaxies with $\dot M_\star\ga 30\, M_\odot$ yr$^{-1}$.

The agreement with all these datasets is substantially improved when basing on the cosmic SFR density computed from the UV+far-IR/sub-mm/radio SFR functions. We also illustrate the contribution to the total density from objects with spheroid-like and disc-like star-formation histories. We remark that at $z\ga 1$ most of the SFR density is contributed by dusty starforming progenitors of local massive quiescent spheroids, while at $z\la 1$ it is contributed both by disc-dominated galaxies and by low-mass spheroids, irregulars, and reactivated massive galaxies.

\subsection{Star-formation history of individual galaxies}\label{sec|SFR_hist}

The second ingredient of our analysis is constituted by deterministic evolutionary tracks for the history of star formation in individual galaxies. The relevant quantity $\dot M_\star(\tau|M_\star,t)$ is the behavior of the SFR as a function of the internal galactic age $\tau$ (i.e., the time since the beginning of significant star formation activity) for a galaxy with relic stellar mass $M_\star$ at cosmological time $t$ (corresponding to redshift $z$).

For high $z\ga 1$ strongly star-forming galaxies (that will turn out to be the progenitors of local dead massive spheroids), we base on the indications emerging from many SED-modeling studies (e.g., Papovich et al. 2011; Smit et al. 2012; Moustakas et al. 2013; Steinhardt et al. 2014; Citro et al. 2016; Cassar\'a et al. 2016); these suggest a slow power-law increase of the SFR $\dot M_\star\propto \tau^\kappa$ with $\kappa\la 1$ over a timescale $\tau_{\rm sphe}\la $ Gyr, then followed by a rapid quenching, at least for massive objects. For the sake of simplicity (cf. Sect.~\ref{sec|continuity}), here we adopt the law
\begin{equation}\label{eq|SFR_time_etg}
\nonumber \dot M_\star(\tau|M_\star,t) = {\kappa+1\over 1-\mathcal{R}}\, {M_\star\over \tau_{\rm sphe}}\,(\tau/\tau_{\rm sphe})^\kappa\, \Theta_{\rm H}\left[\tau\leq \tau_{\rm sphe}\right]~,\\
\end{equation}
where $\Theta(\cdot)$ is the Heaviside step function. The quantity $\mathcal{R}$ is the fraction of mass restituted to the interstellar medium by massive stars, computed in the instantaneous recycling approximation; for a Chabrier IMF and star-formation timescales $\la$ Gyr, $\mathcal{R}\approx 0.4$ applies.

As to the parameters involved in the above expressions, recent observations by \textsl{ALMA} have shown that in high-redshift galaxies the star formation occurred within a compact region $\la$ a few kpcs over timescales $\tau_{\rm sphe}\la 0.5-1$ Gyr at violent rates $\dot M_\star\ga$ a few $10^2\, M_\odot$ yr$^{-1}$ under heavily dust-enshrouded conditions (e.g., Scoville et al. 2014, 2016; Simpson et al. 2015; Ikarashi et al. 2015; Straatman et al. 2015; Spilker et al. 2016; Tadaki et al. 2017). A duration of the main star formation episode $\tau_{\rm sphe}\la 0.5-1$ Gyr in high-redshift dusty star-forming galaxies, which are the candidate progenitors of massive spheroids, is also confirmed by local observations of the $\alpha-$enhancement, i.e., iron underabundance compared to $\alpha$ elements. This occurs because star formation is stopped, presumably by some form of energetic feedback (e.g., due to the central supermassive black hole), before type I$a$ surpernova explosions can pollute the interstellar medium with substantial iron amounts (e.g., Romano et al. 2002; Thomas et al. 2005; Gallazzi et al. 2006; for a review see Renzini 2006). Contrariwise, in low-mass spheroidal galaxies with $M_\star\la 10^{10}\, M_\odot$ data on the age of stellar population and on chemical abundances indicate that star formation has proceeded for longer times, mainly regulated by supernova feedback and stellar winds (see review by Conroy 2013).

\begin{figure}
\epsscale{1}\plotone{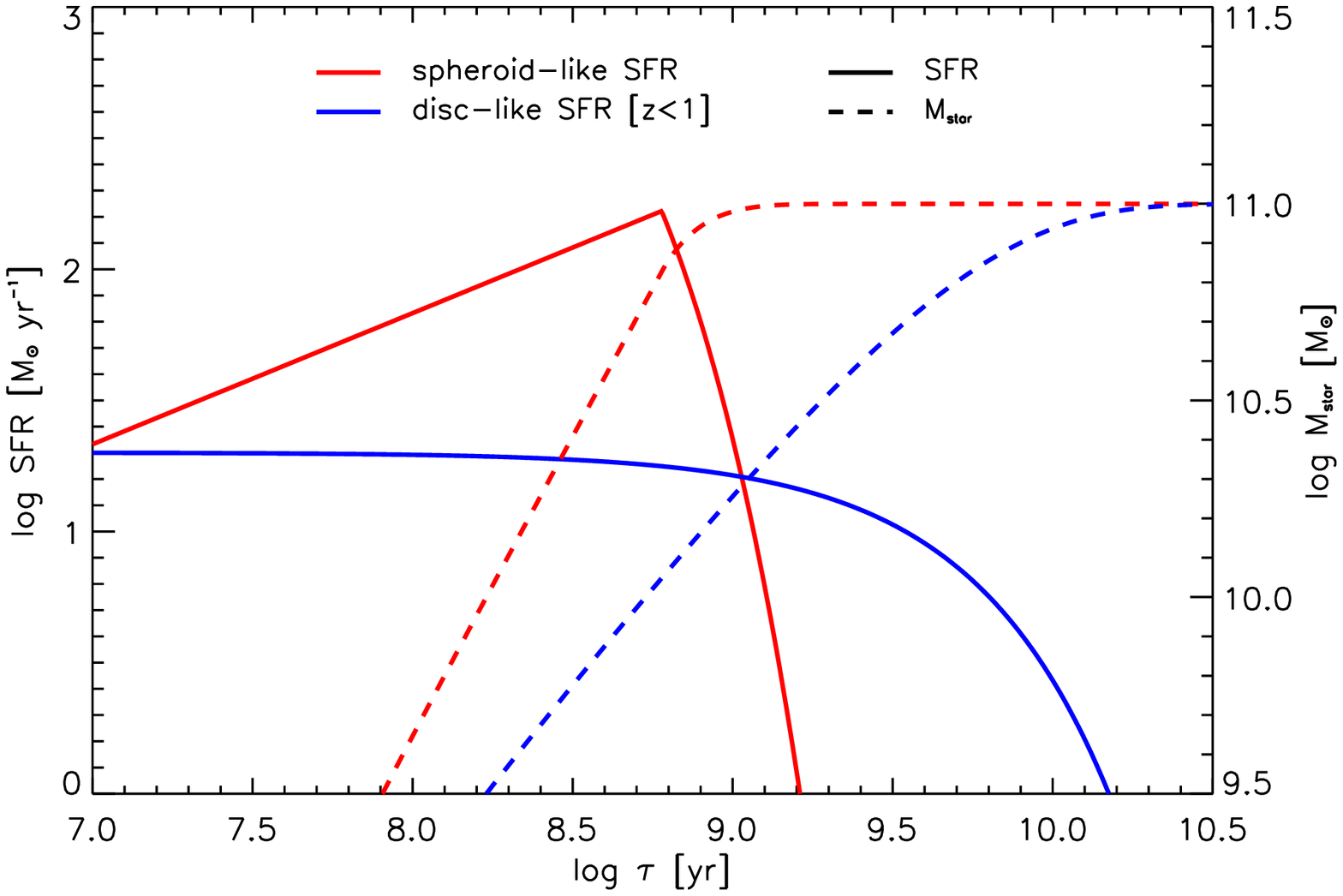}\\\plotone{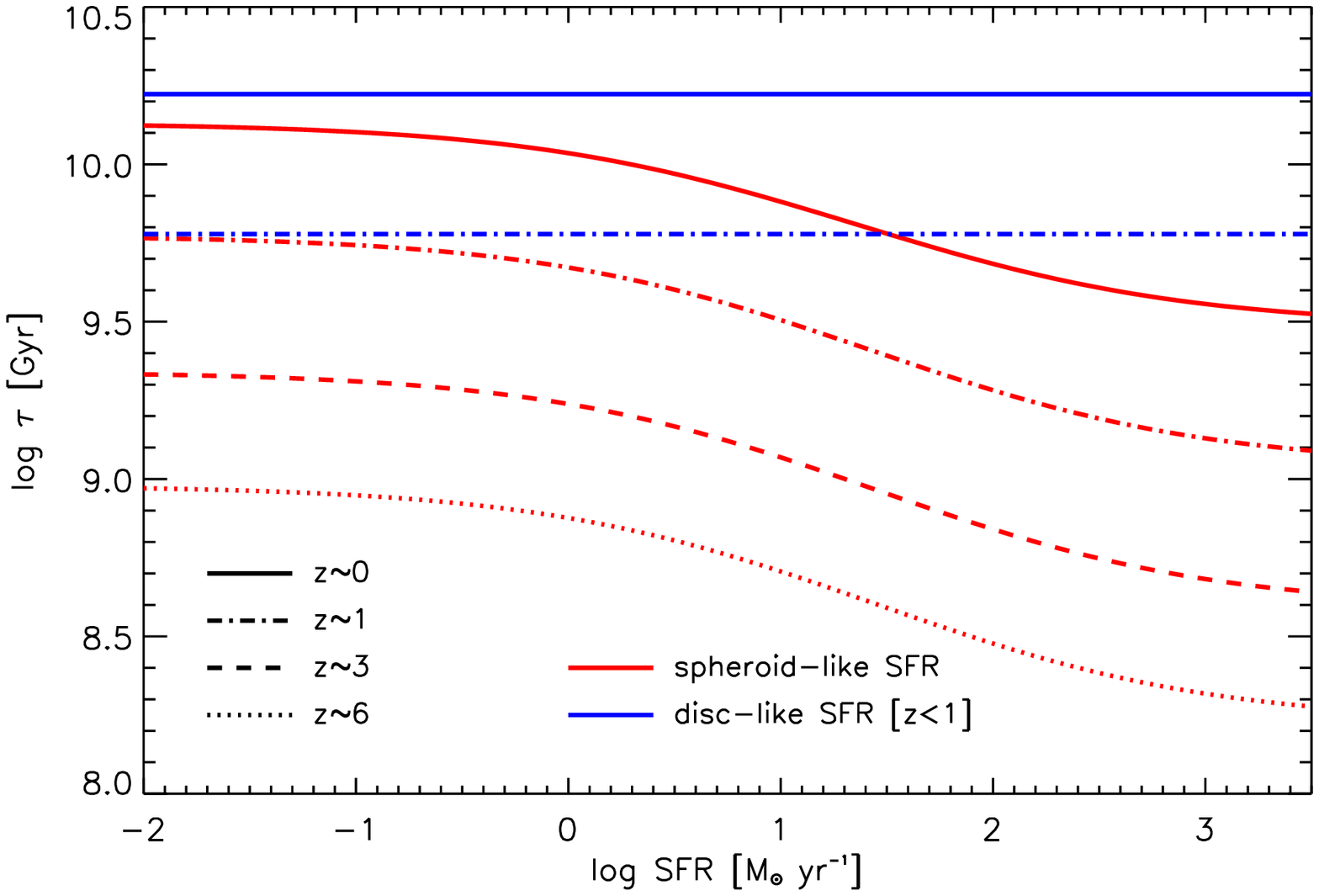}\caption{Top panel: evolution of the SFR (solid lines, left $y$-axis) and of the stellar mass (dashed lines, right $y$-axis) as a function of the galactic age $\tau$ for spheroid-like  (see Eq.~\ref{eq|SFR_time_etg}; red lines) and disc-like (see Eq.~\ref{eq|SFR_time_ltg}; blue lines) star-formation histories; both evolutions refer to a galaxy with final stellar mass $M_\star\approx 10^{11}\, M_\odot$. Bottom panel: The star-formation timescales $\tau_{\rm sphe}$ for spheroid-like (see Eq.~\ref{eq|tau_burst_etg}; red lines) and disc-like objects (only plotted at $z\la 1$, see Eq.~\ref{eq|SFR_time_ltg}; blue lines) as a function of the SFR and at redshift $z\approx 0$ (solid), $1$ (dot-dashed), $3$ (dashed), and $6$ (dotted).}\label{fig|SFR_time}
\end{figure}

On this basis, we parameterize the timescale for the duration of the main SFR episode in objects with spheroid-like star-formation history
as a function of the peak SFR value $\dot M_\star(\tau_{\rm sphe}|M_\star,t)=(\kappa+1)\, M_\star/(1-\mathcal{R})\,\tau_{\rm sphe}$ via the implicit equation
\begin{eqnarray}\label{eq|tau_burst_etg}
\nonumber \tau_{\rm sphe} &=& {\tau_{\rm sphe}^+ +\tau_{\rm sphe}^-\over 2}+{\tau_{\rm sphe}^+ -\tau_{\rm sphe}^-\over 2}\times\\ 
\nonumber\\
&\times& \tanh\left[{\dot M_\star(\tau_{\rm sphe}|M_\star,t)\over 5\, M_\odot\, {\rm yr}^{-1}}\right]~,\\
\nonumber\\
\nonumber \tau_{\rm sphe}^+ &=& 0.6\, {\rm Gyr}\,\left({1+z\over 3}\right)^{-3/2}~,~~~~~~ \tau_{\rm sphe}^- = t_z~;
\end{eqnarray}
this has to be solved on a grid of $M_\star$ and $t$ (or $z$; see details in Sect.~\ref{sec|continuity}). The $\tanh(\cdot)$ function interpolates smoothly between the short timescale $\tau_{\rm sphe}^+\la 1$ Gyr for high star-forming galaxies, and a long timescale $\tau_{\rm sphe}^-\sim t_z$ of the order of the cosmic time for galaxies with low SFRs. Note that in $\tau_{\rm sphe}^+$ the dependence on redshift matches that of the dynamical time $\propto 1/\sqrt{G\,\rho}\propto (1+z)^{-3/2}$, in turn following the increase in average density $\rho\propto (1+z)^3$ of the ambient medium. Our results will be insensitive to the specific shape of the smoothing function. With a similar parameterization, Mancuso et al. (2016b) have been able to reproduce the main sequence of star-forming galaxies at $z\approx 2$. We recall that at $z\la 1$, since the knee of the SFR functions recedes a lot, most of the objects with spheroid-like star-formation histories are characterized by moderate SFR $\dot M_\star\la 10\, M_\odot$ yr$^{-1}$, hence rather long star-formation timescales up to a few Gyrs (cf. Fig.~\ref{fig|SFR_time}, bottom panel).

As to the quenching timescale, the observed fraction of far-IR detected host galaxies in X-ray (e.g., Mullaney et al. 2012; Page et al. 2012; Rosario et al. 2012; Barger et al. 2015; Stanley et al. 2015; Harrison et al. 2016) and optically selected AGNs (e.g., Mor et al. 2012; Wang et al. 2013; Willott et al. 2015; Xu et al. 2015; Netzer et al. 2016; Harris et al. 2016) points toward a SFR abruptly stopping, at least in massive galaxies, after $\tau_{\rm sphe}$ over a short timescale $\la 10^8$ yr due to the action of feedbacks (see Lapi et al. 2014). To avoid introducing an additional parameter, in Eq.~(\ref{eq|SFR_time_etg}) we truncate the SFR abruptly after $\tau_{\rm sphe}^+$; we checked that an exponential quenching over a short timescale $\la \tau_{\rm sphe}^+/\zeta$ with $\zeta\ga$ a few will add an additional, small delay and produce very similar outcomes on the stellar mass function.

On the other hand, in low redshift $z\la 1$ disc-dominated galaxies, it is well known that on average star formation declines exponentially as a function of the galactic age, with a long characteristic timescale of several Gyrs; for example, for our Milky Way it amounts to $\approx 6-7$ Gyrs (see Chiappini et al. 1997; Courteau et al. 2014; Pezzulli \& Fraternali 2016). In view of these classic evidences, we adopt
\begin{eqnarray}\label{eq|SFR_time_ltg}
\nonumber\dot M_\star(\tau|M_\star,t) &=& {1\over 1-\mathcal{R}}\,{M_\star\over \tau_{\rm disc}}\, e^{-\tau/\tau_{\rm disc}}~,\\
\\
\nonumber\tau_{\rm disc} &=& 6\, {\rm Gyr}\,\left({1+z\over 2}\right)^{-3/2}~.
\end{eqnarray}
In Fig.~\ref{fig|SFR_time} (top panel) we illustrate an example of the resulting spheroid-like and disk-like star-formation histories; the relative star-formation timescales as a function of SFR and cosmic time are also shown (bottom panel). For both disc- and spheroid-like histories, we assume a dispersion of $0.25$ dex around the average star-formation timescales; this value is inspired by the scatter observed in the specific SFRs $\dot M_\star/M_\star$ (the inverse of a mass doubling time) of active galaxies at different redshift (e.g., Madau \& Dickinson 2014), and it will turn out to produce the observed dispersion in the resulting star formation main sequence (cf. Sect.~\ref{sec|results} and Fig.~\ref{fig|mainseq}).

We caveat the reader that in the literature there have been attempts to parameterize with a unique shape the average star-formation history of galaxies. A classic way involves the so-called 'delayed exponential' model $\dot M_\star(\tau)\propto \tau^{\kappa}\, e^{-\tau/\tau_{\star}}$, with two parameters $\kappa$ and $\tau_\star$ controlling the early powerlaw rise and the late exponential decline. More recently, analogy with the behavior of the cosmic SFR density (see Gladders et al. 2013) and indications from numerical simulations (see Diemer et al. 2017) have suggested a lognormal shape $\dot M_\star(\tau)\propto e^{-(\ln \tau/\tau_\star)^2/2\sigma_\star^2}/\sqrt{2\pi\sigma_\star^2}\, \tau$ with  the parameters $\tau_\star$ and $\sigma_\star$ controlling peak time and width. Other descriptions with more complex parametric form have also been proposed based on observations (e.g., Leitner \& Kravtsov 2011) or empirical models (e.g., Behroozi et al. 2013; Moster et al. 2013). All these shapes can be useful to describe the star-formation history averaged over the entire population of a galaxy survey; however, chemical and photometric data
require to differentiate between disc-like and spheroid-like star-formation histories, making the parametric models for each class (e.g., see Fig. 2 in Diemer et al. 2017) essentially indistinguishable from our simple adopted shapes. For example, to describe the history of a starforming disc the timescale of the early rise has to be much faster than that of the late decline, to mirror the exponential model of Eq.~(\ref{eq|SFR_time_ltg}); contrariwise, in a massive spheroid progenitor the SFR must be nearly constant and then abruptly quenched, to mirror the power-law truncated model of Eq.~(\ref{eq|SFR_time_etg}).

\section{The continuity equation}\label{sec|continuity}

The continuity equation has been originally devised for connecting the AGN statistics to the demographics of both active and dormant supermassive black holes (Cavaliere et al. 1971; Soltan 1982; Small \& Blandford 1992; Salucci et al. 1999; Yu \& Lu 2004, 2008; Marconi et al. 2004; Merloni \& Heinz 2008; Shankar et al. 2009, 2013). Aversa et al. (2015) have been the first to show that it can be also applied to the stellar component in galaxies, to link the evolution across cosmic times of the SFR function to the stellar mass functions. The continuity equation in integral formulation is written
\begin{widetext}
\begin{equation}\label{eq|continuity}
{{\rm d}N\over {\rm d}\log \dot M_\star}(\dot M_\star,t) = \int{\rm d}\log M_\star~\partial_t \left[{{\rm d}N\over {\rm d}\log M_\star}(M_\star,t)-S(M_\star,t)\right]\, {{\rm d}\tau\over {\rm d}\log \dot M_\star}(\dot M_\star|M_\star,t)~;
\end{equation}
\end{widetext}
here the term on the l.h.s. is the (known) SFR function, while under the integral on the r.h.s. the first factor is the cosmic time derivative of the (unknown) stellar mass function minus a source term due to dry mergers (i.e., adding the whole mass content in stars of merging objects without contributing significantly to in-situ star formation), and the second factor is the overall time spent by a galaxy in a bin of SFR obtained from the star formation history. The interested reader can find in Aversa et al. (2015) an extended discussion of how and under which hypothesis the standard differential form of the continuity equation is recovered.

In general, the continuity equation above is integro-differential and has to be solved numerically. If the source term due to dry merging is negligible (as it turns out to be indeed for $z\ga 1$ according to simulations, see Sect.~\ref{sec|drymerg} for details) and the star formation histories have simple shapes like in Eqs.~(\ref{eq|SFR_time_etg}) and (\ref{eq|SFR_time_ltg}), the continuity equation can be solved analytically along the following lines (see Aversa et al. 2015). First, the time lapses spent by the galaxy in a logarithmic bin of SFR read
\begin{widetext}
\begin{eqnarray}\label{eq|dtau}
\nonumber {{\rm d}\tau\over {\rm d}\log \dot M_\star}(\dot M_\star|M_\star,t) &=& {1\over \kappa\,(\kappa+1)^{1/\kappa}}\, {\dot M_\star^{1/\kappa}\over M_\star^{1/\kappa}}\,\tau_{\rm sphe}^{1+1/\kappa}\, \ln(10)\, \Theta_{\rm H}\left[\dot M_\star\leq {(\kappa+1)\,M_\star\over (1-\mathcal{R})\,\tau_{\rm sphe}}\right]~,\\
& &\\
\nonumber &=& \tau_{\rm disc}\, \ln(10)\, \Theta_{\rm H}\left[\dot M_\star\leq {M_\star\over (1-\mathcal{R})\, \tau_{\rm disc}}\right]~,
\end{eqnarray}
\end{widetext}
for galaxies with spheroid- and disc-like star-formation histories, respectively. In both expressions the Heaviside step function $\Theta_{\rm H}(\cdot)$ specifies the maximum SFR contributing to a given final stellar mass.

Inserting these expressions in the continuity equation Eq.~(\ref{eq|continuity}), differentiating with respect to $\dot M_\star$ and then integrating over cosmic time yield the closed form solutions
\begin{widetext}
\begin{eqnarray}\label{eq|solution}
\nonumber {{\rm d}N(\log M_\star,t)\over {\rm d}\log M_\star} &=& -\kappa\,(1+\kappa)^{1/\kappa}\, M_\star^{1/\kappa}\, \int_{0}^t{\rm d}t'~{\partial_{\ln\dot M_\star}\over
f_{\tau_{\rm sphe}}}\,\left[\dot M_\star^{-1/\kappa}\,
\tau_{\rm sphe}^{-1-1/\kappa}\, {{\rm d}N\over {\rm d}\log \dot M_\star}(\dot M_\star,t')\right]_{\big{|}\dot M_\star = {(1+\kappa)\,M_\star\over (1-\mathcal{R})\,\tau_{\rm sphe}}}~;\\
& &\\
\nonumber &=& -\int_{0}^t{\rm d}t'~{\partial_{\ln\dot M_\star}\over
f_{\tau_{\rm disc}}}\,\left[\tau_{\rm disc}^{-1}\,{{\rm d}N\over {\rm d}\log \dot M_\star}(\dot M_\star,t')\right]_{\big{|}\dot M_\star = {M_\star\over (1-\mathcal{R})\,\tau_{\rm disc}}}~,
\end{eqnarray}
\end{widetext}
again for galaxies with spheroid- and disc-like star-formation histories, respectively; in both expression we have used the shorthand $f_{\tau}\equiv 1+\partial_{\log\dot M_\star}\log\tau$, which is not trivially equal to one when $\tau$ depends explicitly on the SFR (as in Eq.~\ref{eq|tau_burst_etg}). The above equation is numerically solved on a grid in $M_\star$ and $z$ (or cosmic time $t_z$). We use a grid of $100$ equally-spaced points in $\log M_\star [M_\odot]$ within the range $[8,13]$ and a grid of $1000$ equally-spaced points in redshift $z$ within the range $[0,20]$; for optimal interpolation, the SFR functions and the star-formation timescales have been defined on the same grid of redshift and on a grid of $100$ equally-spaced points in $\log \dot M_\star [M_\odot$ yr$^{-1}]$ within the range $[-2,4]$.

\subsection{Dry merging}\label{sec|drymerg}

In presence of mergers, the source $S(M_\star,t)=S_+-S_-$ actually includes the difference between a creation $S_+$ and a destruction $S_-$ term. The former depends on the merger rate of objects with smaller masses into the descendant mass $M_\star$, while the latter depends on the merger rates of the mass $M_\star$ into more massive objects. Given the merger rate ${{\rm d}N_{\rm merg}/{\rm d}\log M_\star\,{\rm d}\mu\,{\rm d}t}$ for the production of a descendant mass $M_\star$ by the merging of two progenitors with (smaller to higher) mass ratio $\mu$, the creation term reads
\begin{equation}\label{eq|creation}
S_+(M_\star,t) = {1\over 2}\, \int_{\mu_{\rm min}}^{1}{\rm d}\mu~{{\rm d}N_{\rm merg}\over {{\rm d}\log M_\star\,\rm d}\mu\,{\rm d}t}(M_\star,\mu,t)
\end{equation}
while the destruction term is written
\begin{eqnarray}\label{eq|destruction}
\nonumber S_-(M_\star,t) &=& {1\over 2} \int_{\mu_{\rm min}}^{1}{\rm d}\mu\,\left[{{\rm d}N_{\rm merg}\over {\rm d}\log M_\star\,{\rm d}\mu\,{\rm d}t}(M_\star(1+\mu)/\mu,\mu,t)+\right.\\
\nonumber \\
&+& \left. {{\rm d}N_{\rm merg}\over {\rm d}\log M_\star\,{\rm d}\mu\,{\rm d}t}(M_\star\,(1+\mu),\mu,t)\right]~.
\end{eqnarray}
In the above $\mu_{\rm min}$ is the minimum progenitors' mass ratio; typically $\mu_{\rm min}=0.3$ includes only 'major mergers', $0.1$ includes major and minor mergers, $\la 0.1$ practically includes all mergers. We take $\mu_{\rm min}=0.01$ in the following.

We base on the outcomes of the \textsl{Illustris} simulations by Rodriguez-Gomez et al. (2015, 2016), who provide a handy fitting function for the merger rate per descendant galaxy
\begin{eqnarray}
&\nonumber& {{\rm d}n_{\rm merg}\over {\rm d}\mu\,{\rm d}t}(M_\star,\mu,t) = A(z)\, \left(M_\star\over 10^{10}\, M_\odot\right)^{\omega(z)}\times\\
\\
&\nonumber &\times \left[1+\left({M_\star\over 2\times 10^{11}\, M_\odot}\right)^{\delta(z)}\right]\, \mu^{\beta(z)+\gamma\,\log(M_\star/10^{10}\, M_\odot)}
\end{eqnarray}
where
\begin{eqnarray}
\nonumber A(z)&=&A_0\,(1+z)^\eta~,~~~~\omega(z)=\omega_0\,(1+z)^{\omega_1}~,\\
\\
\nonumber \beta(z)&=&\beta_0\,(1+z)^{\beta_1},~~~~\delta(z)=\delta_0\,(1+z)^{\delta_1}
\end{eqnarray}
with $A_0\approx 10^{-2.2287}$ Gyr$^{-1}$, $\eta\approx 2.4644$, $\omega_0\approx 0.2241$, $\omega_1\approx-1.1759$, $\beta_0\approx-1.2595$, $\beta_1\approx 0.0611$, $\gamma\approx -0.0477$, $\delta_0\approx 0.7668$, $\delta_1\approx -0.4695$. The above authors have validated this expression against various datasets, including observations of galaxy pairs.
Two remarks are in order here. First, we caveat that at $z\ga 1$ the \textsl{Illustris} simulation does not perfectly reproduce the observed galaxy stellar mass function; however, this does not concern much the merger rates, since as we shall demonstrate the growth in stellar mass at high redshift is mainly dominated by in situ star formation. Second, our results will turn out to be robust against other choices of the merger rate; we checked this by exploiting the galaxy merger rates extracted from the hydrodynamic simulations by Stewart et al. (2009), and the halo merger rates based on the $N-$body simulation by Fakhouri et al. (2010) coupled with empirical relationships connecting halo and stellar mass (e.g., Moster et al. 2013).

Multiplying the merger rates above by the stellar mass function yields the quantity
\begin{eqnarray}
\nonumber {{\rm d}N_{\rm merg}\over {\rm d}\log M_\star\,{\rm d}\mu\,{\rm d}t}(M_\star,\mu,t)&=&{{\rm d}n_{\rm merg}\over {\rm d}\mu\,{\rm d}t}(M_\star,\mu,t)\times\\ 
\\
\nonumber &\times& {{\rm d}N\over {\rm d}\log M_\star\,{\rm d}t}(M_\star,t)
\end{eqnarray}
entering the source terms in Eqs.~(\ref{eq|creation}) and (\ref{eq|destruction}). As discussed by Rodriguez-Gomez et al. (2015, 2016) the above expression also takes into account stellar mass stripping from satellites prior to dry mergers. Plainly, when merging is introduced, the continuity equation becomes fully integro-differential and must be solved numerically. We have computed the full solution and found that, to a good approximation, one can solve the problem iteratively, using Eq.~(\ref{eq|solution}) as the zero-th order solution and then updating it with the correction due to the merging terms.

\subsection{Cosmic stellar mass density and Soltan argument}\label{sec|massden}

Once the redshift-dependent stellar mass function is known, the cosmic stellar mass density is obtained as
\begin{equation}
\rho_{M_\star}(t) \equiv \int{\rm d}\log M_\star\, M_\star\, {{\rm d}N\over {\rm d}\log M_\star}(M_\star,t)~,
\end{equation}
where the integration is typically performed over stellar masses above $10^8\, M_\odot$ for fair comparison with observational determinations (see discussion by Madau \& Dickinson 2014).

Interestingly, if $\tau_{\rm sphe,disc}$ is independent of, or only weakly dependent on $\dot M_\star$, a Soltan (1982) argument holds for the stellar content of galaxies (see Aversa et al. 2015); classically, this connects the cosmic luminosity density to the relic mass density of a population via an average conversion efficiency. In the present context, the Soltan argument can be easily found by multiplying both sides of Eq.~(\ref{eq|continuity}) by $\dot M_\star$ and integrating over it and over cosmic time, to obtain
\begin{equation}\label{eq|Mstar_cosm}
\rho_{M_\star} = (1-\mathcal{R})\,\int_0^t {\rm d}t'\, \int{\rm d}\log \dot M_\star\, \dot M_\star\,{{\rm d}N\over {\rm d}\log \dot M_\star}(\dot M_\star,t')~.
\end{equation}
We highlight that here the IMF-dependent factor $1-\mathcal{R}$ plays the role of the radiative efficiency in the classic Soltan argument for black holes. Note that for conventional IMFs most of the stellar mass in galaxies resides in stars with mass $\la 1\, M_\odot$; since these stars emit most of their luminosity in the near-IR, the galaxy stellar mass $M_\star$ can be inferred by the near-IR luminosity functions. On the other hand, the SFR function is determined from UV and far-IR/sub-mm/radio observations, as discussed in Sect.~\ref{sec|basics}. Thus in principle accurate determinations of the SFR and stellar mass functions in these independent manners could be exploited via the Soltan argument above to constrain the average galaxy IMF at different redshifts. In practice, however, the dependence on the IMF is weak, and current observational uncertainties do not allow to fulfill this program now.

\subsection{Star formation efficiency}\label{sec|SFE}

We now connect the stellar mass function to the underlying, gravitationally dominant DM component, with the aim of deriving the star formation efficiency $f_\star\equiv M_\star/f_{\rm b}\, M_{\rm H}$. This represents the fraction of the baryonic mass $f_{\rm b}\, M_{\rm H}\approx 0.16\, M_{\rm H}$ initially associated to a DM halo of mass $M_{\rm H}$ that has been eventually converted into stars. To this purpose, we exploit the abundance matching technique, a standard way of deriving a monotonic relationship between galaxy and halo properties by matching the corresponding integrated number densities (e.g., Vale \& Ostriker 2004; Shankar et al. 2006; Moster et al. 2013; Behroozi et al. 2013).

\begin{figure*}
\epsscale{1}\plotone{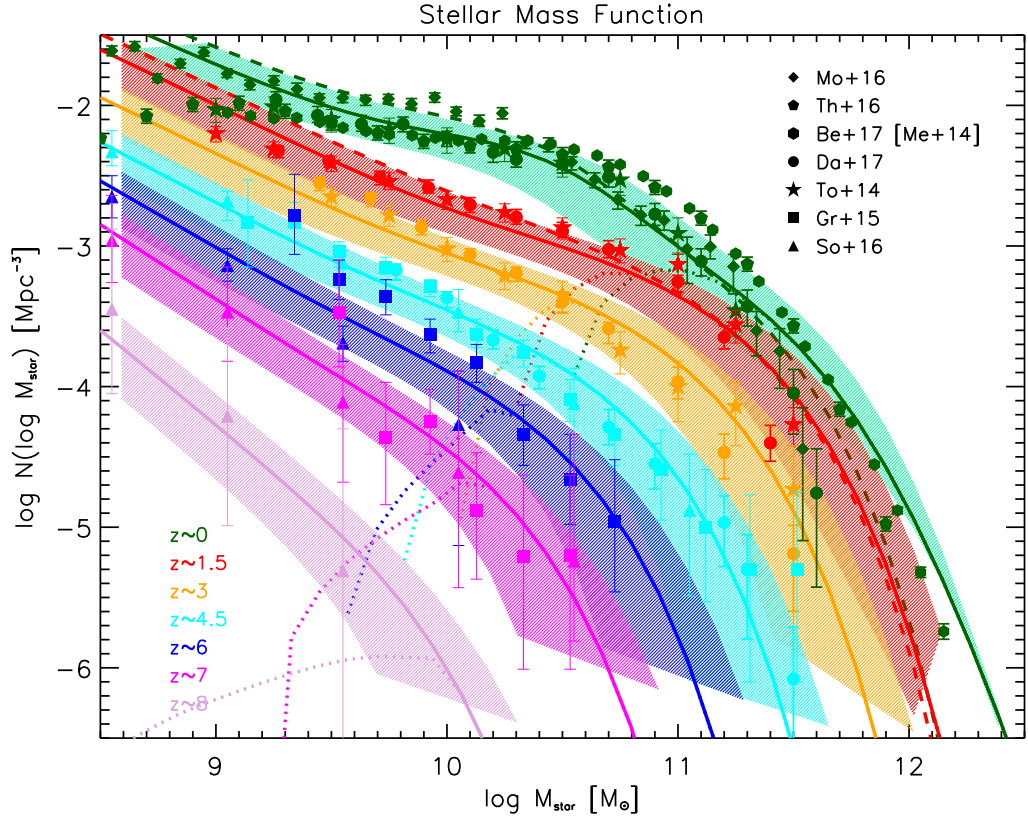}\caption{The stellar mass function at redshifts $z=0$ (green), $1.5$ (red), $3$ (orange), $4.5$ (cyan), $6$ (blue), $7$ (magenta), and $8$ (lilac), determined via the continuity equation including both in situ star formation and dry mergers (solid lines); shaded areas represent the $1\sigma$ uncertainty resulting from the scatter in star formation timescales (see Sect.~\ref{sec|basics} for details). Dashed lines (actually superimposed to the solid ones at $z\ga 1.5$) refer to the outcome of including only in-situ star formation, and dotted lines show the contribution to the stellar mass function from galaxies with spheroid-like star-formation histories that featured SFR exceeding $100\, M_\odot$ yr$^{-1}$. Data are from Moffett et al. (2016; diamonds), Thanjavur et al. (2016; pentagons), Bernardi et al. (2017, based on the $M/L$ ratios by Mendel et al. 2014; hexagons), Davidzon et al. (circles), Tomczak et al. (2014; stars), Grazian et al. (2015; squares), and Song et al. (2016; triangles).}\label{fig|Mstar_func_TOT}
\end{figure*}

For fair comparison with the determination of the star formation efficiency $f_\star\equiv M_\star/f_{\rm b}\, \langle M_{\rm H}\rangle$ via weak gravitational lensing (e.g., Velander et al. 2014; Hudson et al. 2015; Mandelbaum et al. 2016) and galaxy kinematics (e.g., More et al. 2011; Wojtak \& Mamon 2013), that are based on galaxy samples selected by stellar mass, we aim at deriving the average halo mass $\langle M_{\rm H}\rangle (M_\star,z)$ associated to a given $M_\star$. In the abundance matching formalism, this relationship is obtained via the equation (see Aversa et al. 2015 for details)
\begin{eqnarray}\label{eq|abmatch}
&\nonumber& \int_{\log \langle M_{\rm H}\rangle (M_\star,z)}^\infty{\rm d}\log M_{\rm H}'\, {{\rm d}N\over {\rm d}\log M_{\rm H}}(M_{\rm H}',z) = \\
\\
\nonumber &=&\int_{-\infty}^{+\infty}{\rm d}\log M_\star'\, {{\rm d}N\over {\rm d}\log M_\star}(M_\star',z)\,{1\over 2}\, {\rm erfc}\left\{{\log[M_\star/M_\star']\over \sqrt{2}\,\sigma_{\log M_{\rm H}}}\right\}~,
\end{eqnarray}
holding when a lognormal distribution of $M_{\rm H}$ at given $M_\star$
with dispersion $\sigma_{\log M_{\rm H}}$ is assumed.

We follow previous studies based on various semi-empirical methods of galaxy and halo connection (see Rodriguez-Puebla et al. 2015, their Fig.~10) and adopt $\sigma_{\log M_{\rm H}}\approx \max\left[0.05,0.05+0.15\,(\log M_\star [M_\odot]-10)\right]$ for $\log M_\star [M_\odot]$ within the range $[8.5,12.5]$. In Eq.~(\ref{eq|abmatch}) the quantity ${\rm d}N/{\rm d}\log M_{\rm H}$ is usually taken as the halo mass function from $N-$body simulations (e.g., Tinker et al. 2008; Watson et al. 2013; Bocquet et al. 2016; Comparat et al. 2017), that includes galaxy groups and clusters. This is particularly suitable when comparing with observational determinations of the star formation efficiency based on weak gravitational lensing (see references above), that integrate all the DM mass along the line of sight, including that associated to the surrounding galaxy environment.

However, in order to infer the star formation efficiency of individual galaxies, and not of a galaxy system like a group or a cluster, it would be more appropriate to use the galaxy halo mass function, i.e., the mass function of halos hosting one individual galaxy. This can be built up from the overall halo mass function by adding to it the contribution of subhalos,
and by probabilistically removing from it the contribution of halos corresponding to galaxy systems via halo occupation distribution modeling. We defer the reader to Appendix A of Aversa et al. (2015) for details on such a procedure.

\begin{figure*}
\epsscale{1}\plotone{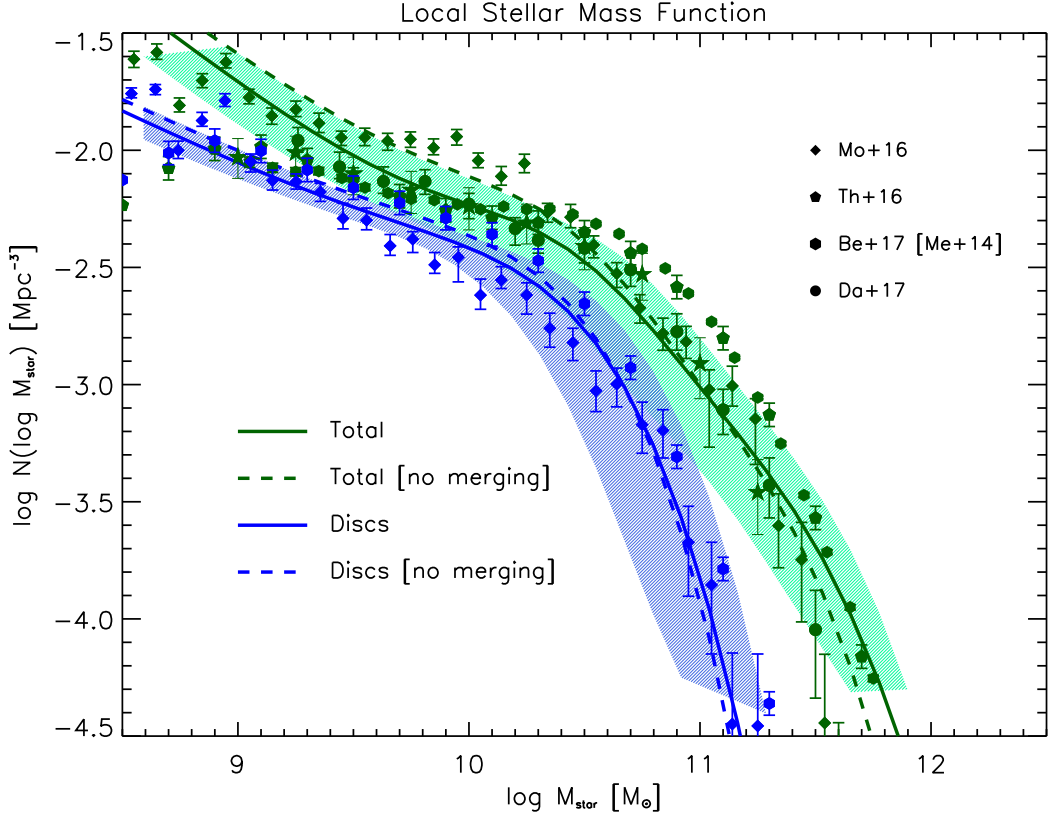}\caption{The stellar mass function at redshifts $z=0$. Green solid line and shaded area refer to the total galaxy population (average and $1\sigma$ uncertainty), while blue solid lines and shaded area refer to galaxies with disc-like star-formation history. The dashed lines highlight the outcomes without including dry mergers (see Sect.~\ref{sec|drymerg} for details). Data are from Moffett et al. (2016; diamonds), Thanjavur et al. (2016; pentagons), Bernardi et al. (2017, based on the $M/L$ ratios by by Mendel et al. 2014; hexagons), and Davidzon et al. (circles).}\label{fig|Mstar_func_z0}
\end{figure*}

\subsection{Galaxy main sequence}\label{sec|MS}

The vast majority of galaxies is endowed with stellar masses strongly correlated to the ongoing SFR, in the way of an almost linear relationship dubbed 'Main Sequence', with a normalization steadily increasing as a function of redshift, and with a limited scatter around 0.25 dex (see Daddi et al. 2007; Rodighiero et al. 2011, 2015; Speagle et al. 2014; Whitaker et al. 2014; Renzini \& Peng 2015; Salmon et al. 2015; Tasca et al. 2015; Kurczynski et al. 2016; Tomczak et al. 2016; Bourne et al. 2017; Dunlop et al. 2017; Schreiber et al. 2017).

We exploit the abundance matching between the SFR functions (Fig.~\ref{fig|SFR_func}) and the stellar mass functions (Fig.~\ref{fig|Mstar_func_TOT}) self-consistently derived from the continuity equation (cf. Eq.~\ref{eq|continuity}) to compute the average SFR $\langle \dot M_\star\rangle(M_\star,z)$ associated to a given stellar mass $M_\star$. This reads
\begin{eqnarray}\label{eq|mainseq}
&\nonumber& \int_{\log \langle \dot M_\star\rangle (M_\star,z)}^\infty{\rm d}\log \dot M_\star'\, {{\rm d}N\over {\rm d}\log \dot M_\star}(\dot M_\star',z) =\\
\\
\nonumber &=& \int_{-\infty}^{+\infty}{\rm d}\log M_\star'\, {{\rm d}N\over {\rm d}\log M_\star}(M_\star',z)\,{1\over 2}\, {\rm erfc}\left\{{\log[M_\star/M_\star']\over \sqrt{2}\,\sigma_{\log \dot M_\star}}\right\}~,
\end{eqnarray}
holding when a lognormal distribution of $\dot M_\star$ at given $M_\star$ with dispersion $\sigma_{\log \dot M_\star}\approx 0.15$ dex is adopted (see Aversa et al. 2015).

The comparison of the resulting main sequence with the observational data will actually constitute an additional constraint on the assumed star formation histories for individual galaxies (see Eqs.~\ref{eq|SFR_time_etg} and \ref{eq|SFR_time_ltg}), on the star formation timescales and the associated scatter, and on the robustness of our results to other aside assumptions discussed in previous sections.

\section{Results}\label{sec|results}

In Fig.~\ref{fig|Mstar_func_TOT} we present the stellar mass function at different redshifts obtained via the continuity equation, including both in situ star formation and (dry) mergers. We highlight the average result as solid lines, and the $1\sigma$ dispersion expected from the scatter in the star formation timescales and merging histories as shaded areas (see Sect.~\ref{sec|basics} for details). We compare our results to recent observational data (Moffett et al. 2016; Thanjavur et al. 2016; Bernardi et al. 2017; Davidzon et al. 2017; Tomczak et al. 2014; Grazian et al. 2015; Song et al. 2016), finding an excellent agreement.

We stress that in-situ star formation within galaxies dominates over dry mergers in building up the stellar mass function at high redshifts, all the way down to $z\sim 1$, while at lower redshifts $z\la 1$ dry mergers can contribute appreciably to the stellar mass growth.  This is highlighted on comparing the solid and dashed lines in Fig.~\ref{fig|Mstar_func_TOT}, which illustrate the mass function at different redshifts when including or not dry mergers, respectively (actually the two sets of curves are superimposed for $z\ga 1.5$). The effect of dry mergers on the stellar mass function is twofold: the number of low mass galaxies is decreased appreciably because of the merging into larger units, whereas the high-mass end of the stellar mass function is boosted toward larger masses because of mass additions from smaller objects; dry mergers mainly affect the most massive galaxies that are typically dominated by the spheroidal component. Such a picture is in agreement with what is recently emerging from state-of-the-art numerical simulations (see Schaye et al. 2015; Rodriguez-Gomez et al. 2016), semiempirical models (see Behroozi et al. 2013) and analysis of observations based on density-matching arguments (see Hill et al. 2017).

\begin{figure*}
\epsscale{1}\plotone{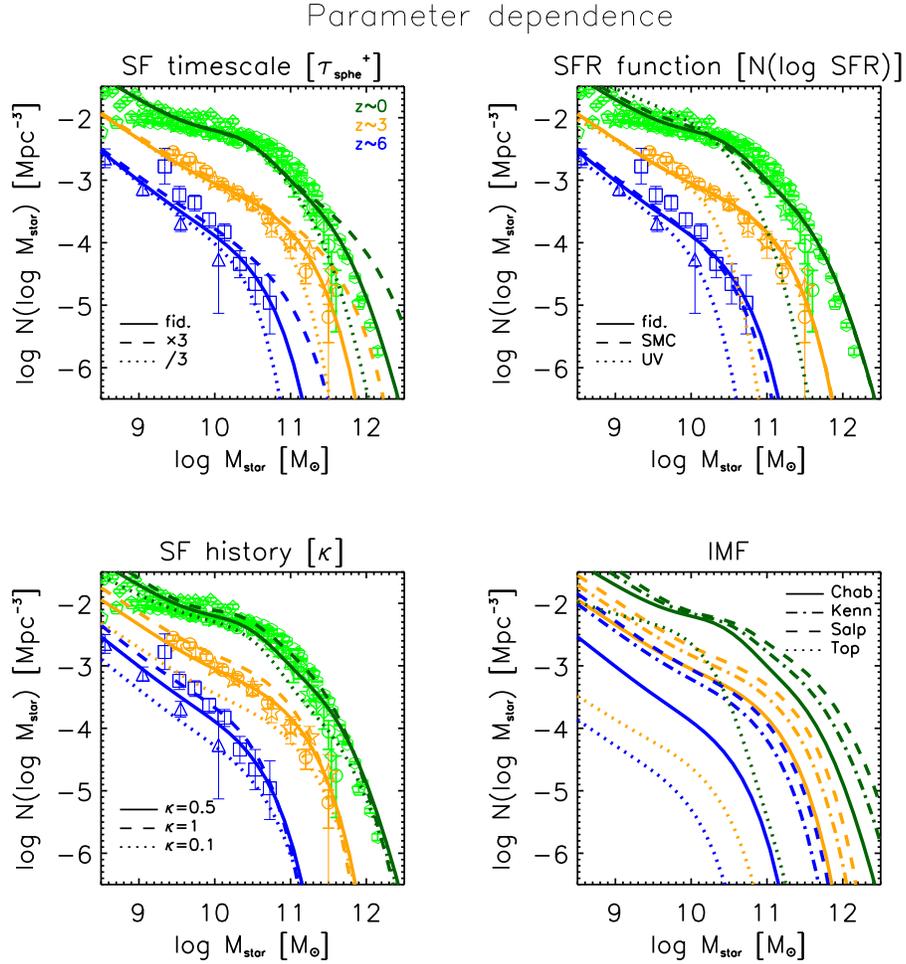}\caption{Parameter dependence of the stellar mass function at redshifts $z=0$ (green), $3$ (orange), and $6$ (blue); data as in Fig.~\ref{fig|Mstar_func_TOT}. Top left panel: dependence on the star-formation timescale $\tau_{\rm sphe}^+$; solid lines refer to the fiducial value in Eq.~(\ref{eq|tau_burst_etg}), dashed lines to a value three times higher and dotted lines to a value three times lower. Bottom left panel: dependence on the parameter $\kappa$ in the star formation history of Eq.~(\ref{eq|SFR_time_etg}); solid lines refer to the fiducial value $\kappa=0.5$, dashed lines to $\kappa=1$, and dotted lines to $\kappa=0.1$. Top right panel: dependence on the input SFR function; solid lines refer the fiducial one represented in Fig.~\ref{fig|SFR_func}, dashed lines to that derived assuming an SMC (in place of the Calzetti) extinction law , dotted line to that inferred from pure UV-dust corrected data (i.e., neglecting far-IR/sub-mm/radio data). Bottom right panel: dependence on the IMF (data are not plotted for clarity); solid lines refer to the fiducial Chabrier IMF, dot-dashed lines to the Kennicutt (1983) IMF, dashed lines to the Salpeter (1995) IMF, and dotted lines to a top-heavy IMF (as in Lacey et al. 2010).}\label{fig|MF_complot}
\end{figure*}

The dotted lines show the contribution to the stellar mass function from galaxies with spheroid-like star-formation history that featured SFRs $\dot M_\star\ga 100\, M_\odot$ yr$^{-1}$; this corresponds to the limiting value currently sampled in wide-area far-IR surveys out to $z\la 4$ (e.g., the \textsl{Herschel}-ATLAS, see Lapi et al. 2011). It is seen that the descendants of these galaxies populate the high-mass end of the local stellar mass function, and thus are mainly present-day massive dead spheroids (e.g., Moffett et al. 2016). This demonstrates on a statistical basis that strongly starforming galaxies observed in the far-IR/(sub-)mm band constitute the progenitors of massive spheroids. By the same token, we stress that to test at $z\ga 4$ the outcomes of the continuity equation, and better constrain the input SFR functions and the parameters of the star-formation history for spheroid progenitors, it will be extremely relevant to improve the accuracy in the determination of the stellar mass function at the high-mass end for $M_\star\ga$ a few $10^{10}\, M_\odot$ out to $z\la 6$ via wide-area near-IR surveys.

\begin{figure*}
\epsscale{1}\plotone{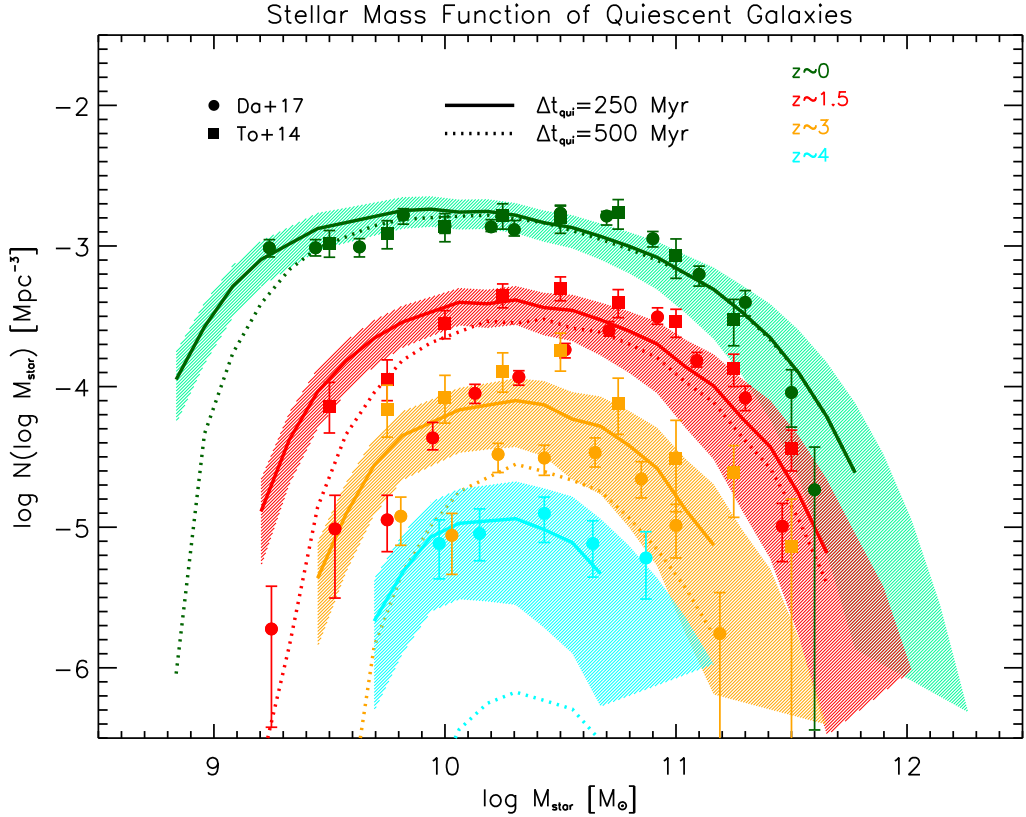}\caption{The stellar mass function of quiescent galaxies at redshifts $z=0$ (green), $1.5$ (red), $3$ (orange), and $4$ (cyan). Solid lines refer to a quiescence timescale of $250$ Myr, and dotted lines to $500$ Myr (see Sect.~\ref{sec|results} for details). Data are from Davidzon et al. (2017; circles) and Tomczak et al. (2014; squares).}\label{fig|Mstar_func_QUI}
\end{figure*}

In Fig.~\ref{fig|Mstar_func_z0} we focus on the stellar mass function of galaxies with disc-like star-formation histories at $z\approx 0$. Our result is in excellent agreement with the observed stellar mass function of disc-dominated galaxies from decomposed data (Moffett et al. 2016; Bernardi et al. 2017); thus we find a good correspondence between objects populating the UV-inferred SFR function, to which we assigned disc-like star formation histories, and galaxies with observed disc-dominated morphology in the stellar mass function. We highlight that discs contribute considerably to the total stellar mass function for stellar masses $M_\star\la$ a few $10^{10}\, M_\odot$, and that the effects of mergers on their stellar mass function are negligible. It is seen from Eq.~(\ref{eq|solution}) that the observed steepness for $M_\star\ga$ a few $10^{10}\, M_\odot$ in the local stellar mass function of disc-dominated galaxies mirrors that in the input UV-inferred SFR functions at $z\la 1$. We caveat that assigning a disc-like star-formation history (with long star-formation timescales) even to objects populating the UV+far-IR/sub-mm/radio SFR functions at $z\la 1$ would considerably overproduce the number of massive discs; this is because the UV+far-IR/sub-mm/radio functions are much higher than the UV-inferred ones at given SFR. In fact, the UV+far-IR/sub-mm/radio SFR functions at $z\la 1$  must be populated by objects with spheroid-like star-formation history; the continuity equation shows these star formation events to change little the total stellar mass function at $z\sim 0$ with respect to that at $z\sim 1$, mildly affecting the number density of galaxies with $M_\star\ga 10^{10}\, M_\odot$.

In Fig.~\ref{fig|MF_complot} we show how our resulting stellar mass
function depend on the input SFR function and on the parameters of the star formation history. To highlight such dependencies in simple terms it is convenient to assume a piecewise powerlaw shape of the SFR function ${\rm d}N/{\rm d}\log \dot M_\star\propto \dot M_\star^{-\chi}$, with $\chi\la 1$ at the faint and $\chi>1$ at the bright end. Then it is easily seen from Eq.~(\ref{eq|continuity}) that the resulting stellar mass function (in absence of mergers) behaves as
\begin{eqnarray}
\nonumber {{\rm d}N\over {\rm d}\log M_\star}&\propto & {1+\kappa\,\chi\over (1+\kappa)^{\chi}}\, (1-\mathcal{R})^{\chi+1/\kappa}\,M_\star^{-\chi}\, \tau_{\rm sphe}^{\chi-1}\\
&& \\
\nonumber &\propto& \chi\, (1-\mathcal{R})^\chi\, M_\star^{-\chi}\,\tau_{\rm disc}^{\chi-1}~,
\end{eqnarray}
for galaxies with spheroid-like and disc-like star formation histories, respectively. Thus, the stellar mass function features an almost direct dependence on the star-formation timescales $\tau_{\rm sphe,disc}$ at the high-mass end, which is mostly contributed by high SFRs where $\chi> 1$; on the other hand, the dependence is inverse but mild at the low-mass end, mainly contributed by low-SFR galaxies with $\chi\la 1$; note, however, that the value of the SFR where $\chi$ appreciably exceeds unity is much lower for the UV-inferred than for the UV+far-IR/sub-mm/radio SFR functions. The dependence on the parameter $\kappa$ entering the star-formation history $\dot M_\star(\tau)\propto \tau^\kappa$ is mild, direct at the low-mass and inverse at the high-mass end. The dependence on the IMF is encapsulated in the restituted fraction $1 - \mathcal{R}$, and in the factor used to convert the observed UV+far-IR/sub-mm/radio luminosity function into the SFR function; e.g., passing from the Chabrier to the Salpeter (1955) IMF, the high mass end of the stellar mass function is increased somewhat, while a strong suppression is originated when basing on a top-heavy IMF (e.g., Lacey et al. 2010). Finally, adopting a SMC extinction law in place of the Calzetti for the determination of the input SFR function amounts to alter somewhat the exponent $\chi$ and thus changes little the final outcome on the stellar mass function; on the other hand, adopting the UV-inferred SFR function at any redshift (i.e., neglecting far-IR/sub-mm/radio data) would imply to strongly underestimate the stellar mass function for large stellar masses (see discussion by Mancuso et al. 2016a,b) that are indeed built up in dusty star-forming galaxies with violent SFRs.

\begin{figure*}
\epsscale{1}\plotone{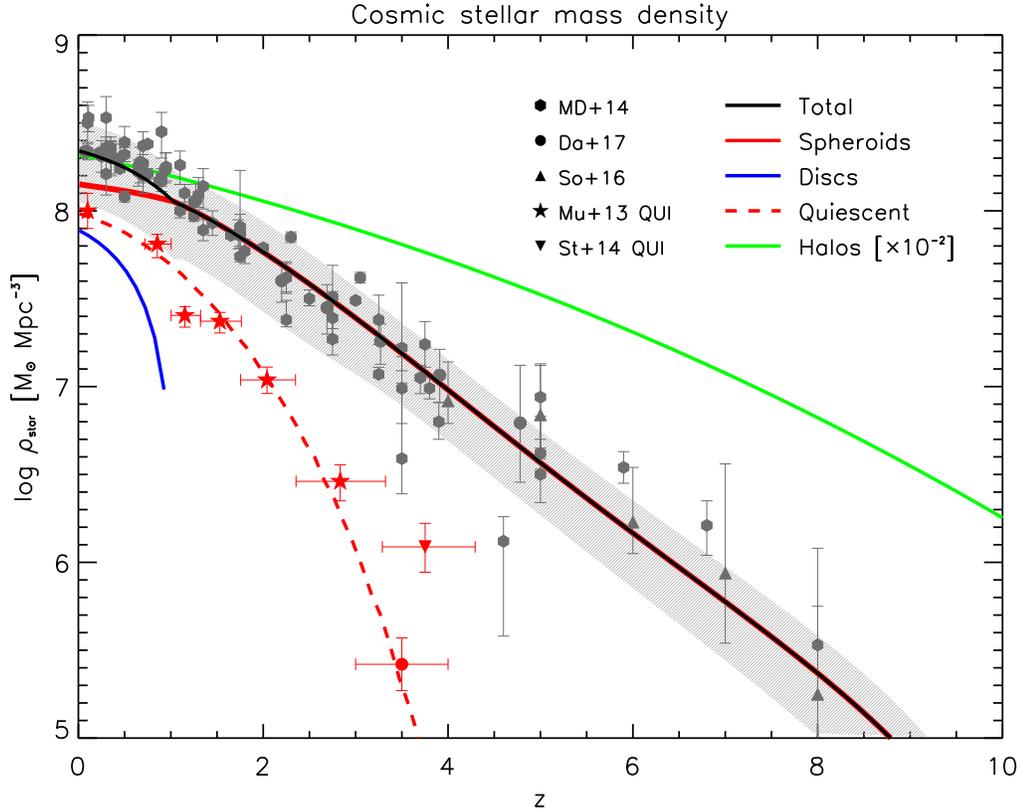}\caption{The cosmic stellar mass density as a function of redshift. Black solid line refers to the total galaxy population, blue solid line to galaxies with disc-like star-formation histories, red solid line to galaxies with spheroid-like star-formation histories, red dashed line to quiescent galaxies (quiescence timescale of $250$ Myr has been adopted). For reference, green solid line is the mass density in dark matter halos, scaled down by a factor of $10^{-2}$. Data are from Madau \& Dickinson (2014), Davidzon et al. (2017; circles), Song et al. (2016; triangles), Muzzin et al. (2013, for quiescent; stars), and Straatman et al. (2014, for quiescent; inverse triangles).}\label{fig|Mstar_cosm}
\end{figure*}

In Fig.~\ref{fig|Mstar_func_QUI} we focus on the stellar mass function of quiescent (passively evolving) galaxies; these systems have been increasingly observed with appreciable number density out to high redshift $z\la 4$ after selection via color-color diagrams in deep near-IR surveys (see Tomczak et al. 2014; Davidzon et al. 2017; Lonoce et al 2017; Glazebrook et al. 2017). Typically, these selections tend to pick up galaxies that have been quenched since, and then passively evolving over, a quiescence time interval $\Delta t_{\rm qui}\sim 250-500$ Myr. Thus we compute the associated stellar mass function from the continuity equation by replacing the upper limit of integration in Eq.~(\ref{eq|continuity}) with $\tau_{\rm sphe}-\Delta t_{\rm qui}$. The result for two different values of $\Delta t_{\rm qui}\approx 250$ and $500$ Myr encompasses very well with the observational determinations out to $z\la 4$. Plainly, higher quiescence time $\Delta t_{\rm qui}$ imply a lower mass function, especially toward higher redshift where the cosmic time is smaller and progressively closer to $\tau_{\rm sphe}$. The decrease of the mass function at the low mass end is due to the fact that small galaxies are still actively forming stars, since they feature longer star formation timescales; as a consequence, the fraction of galaxies in passive evolution decreases rapidly with stellar mass. The downturn shifts toward larger masses toward higher $z$, passing from $10^{10}$ to a few $10^{10}\, M_\odot$ from $z\approx 0$ to $z\ga 3$.

\begin{figure*}
\epsscale{1}\plotone{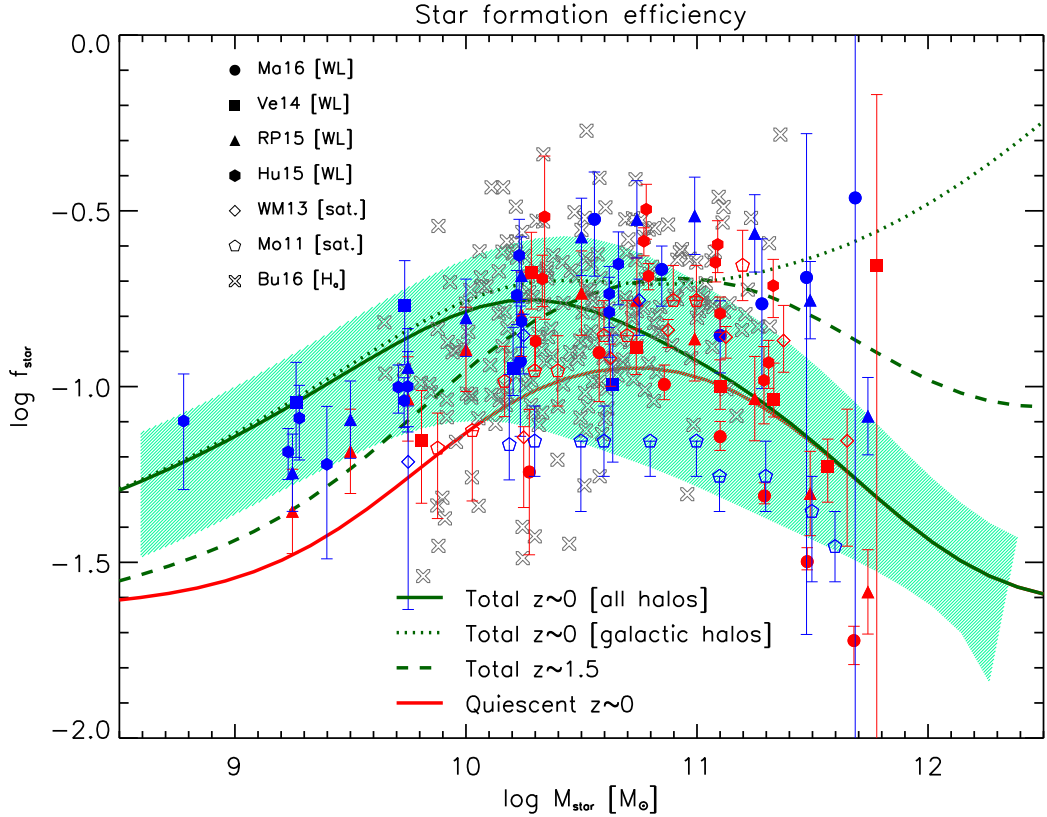}\caption{The star-formation efficiency $f_   \star\equiv M_\star/f_{\rm b}\, \langle M_{\rm H}\rangle$ as a function of the stellar mass $M_\star$, derived from the abundance matching technique (see Sect.~\ref{sec|SFE} for details). Green solid line and shaded area are the average result and its associated $1\sigma$ dispersion at $z\sim 0$ when matching the local stellar mass function to the overall halo mass function; dotted green line is the result at $z\sim 0$ when matching to the galactic halo mass function; dashed green line is the result at $z\sim 1.5$. Solid red line refers to quiescent galaxies at $z\sim 0$. Weak lensing data are from Mandelbaum et al. (2016; circles), Velander et al. (2014; squares), Rodriguez-Puebla et al. (2015; triangles), and Hudson et al. (2015; hexagons); satellite kinematic data are from Wojtak \& Mamon (2013; diamonds) and More et al. (2011; pentagons); $H_\alpha$ data for galaxies at $z\sim 0.8-2.5$ are from Burkert et al. (2016; crosses). Blue symbols are for disc-dominated galaxies and red symbols for spheroids.}\label{fig|fstar}
\end{figure*}

In Fig.~\ref{fig|Mstar_cosm} we show the cosmic stellar mass density, obtained according to Eq.~(\ref{eq|Mstar_cosm}). Our result from integrating the overall stellar mass function from the continuity equation is compared with the data collection by Madau \& Dickinson (2014) and with the recent estimates by Song et al. (2016) and Davidzon et al. (2017) at high redshift. The agreement between our results and the data is remarkably good. We also highlight the contribution to the total stellar mass density from galaxies with disc-like and spheroid-like star-formation histories; the latter dominates the overall mass density at any redshift, though at $z\la 1$ discs brings an appreciable contribution around $40\%$. Note that the fraction of galaxies that are quiescent (here we use $\Delta t_{\rm qui}\approx 250$ Myr, see above discussion), which are basically massive spheroids, constitute only a fraction $50\%$ of the total mass density (contributed also by small spheroids/irregulars and discs that are still active) in the local Universe and rapidly declines to values $\la 10\%$ at higher redshift $z\ga 2$. The overall shape agrees well with the estimates by Muzzin et al. (2013), Straatman et al. (2014) and Davidzon et al. (2017).

For reference, in the figure we also report the mass density (scaled down by a factor $10^{-2}$) of galaxy halos with mass $M_{\rm H}\ga 10^{8.5}\, M_\odot$, the minimal threshold for efficient star formation required to solve the missing  satellite problem (see Boylan-Kolchin et al. 2014; Wetzel et al. 2016; Lapi et al. 2017). The evolution in halos and in the stellar mass content of galaxies differs both in shape and in normalization; these differences stem from: the inefficiency of galaxy formation due to feedback processes (e.g., supernovae, stellar winds, active galactic nuclei); the decrease at high $z$ in the number density of massive halos, that are the hosts of the most massive galaxies; the inability to grow massive galaxies at high redshift since the growth timescales become comparable to the age of the Universe.

\begin{figure*}
\epsscale{1}\plotone{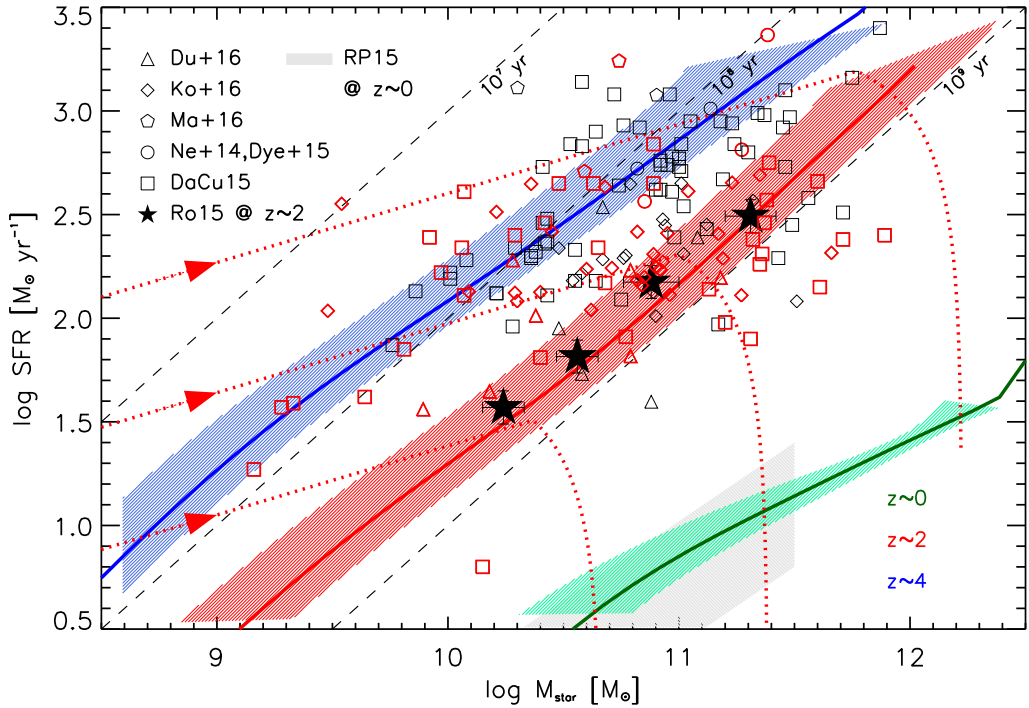}\caption{The main sequence of star-forming galaxies. The colored lines are the outcomes (with shaded areas showing the $1\sigma$ uncertainty) of matching the SFR function and the stellar mass function from the continuity equation (see Sect.~\ref{sec|MS} for details), at redshifts $z\approx 0$ (green), $2$ (red), and $4$ (blue). The red dotted lines show at $z\approx 2$ three evolutionary tracks (forward time direction indicated by arrows) for individual galaxies with peak values of the SFR around $\dot M_\star\approx 20$, $200$, and $2000\, M_\odot$ yr$^{-1}$. The black dashed lines highlight galaxy ages $M_\star/\dot M_\star\approx 10^7$, $10^8$ and $10^9$ yr as labeled.  The black filled stars are the observational determinations of the main sequence at $z\approx 2$ based on statistics of large mass-selected samples by Rodighiero et al. (2015). The other symbols (error bars omitted for clarity) refer to far-IR data for individual objects at $z\sim 1-4$ (those in the range $z\sim 1.5-2.5$ are marked in red and the others in black) by Dunlop et al. (2016; triangles), Koprowski et al. (2016; diamonds), Ma et al. (2015b; pentagons), Negrello et al. (2014) plus Dye et al. (2015; circles), and da Cunha et al. (2015; squares). The grey shaded area is the observational estimate at $z\approx 0$ by Renzini \& Peng (2015).}\label{fig|mainseq}
\end{figure*}

In Fig.~\ref{fig|fstar} we show the star-formation efficiency, computed according to Eq.~(\ref{eq|abmatch}). The green solid line and shaded area illustrate the outcome when matching the total stellar mass function from the continuity equation to the overall halo mass function at $z\approx 0$; note that the shaded area takes into account the uncertainty in the determination of the local stellar mass function from the continuity equation and that arising from the rather flat shape of the average $\langle M_{\rm H}\rangle (M_\star)$ correlation at the high mass end. Our result is compared with the local data for early and late type galaxies by various authors, determined via weak lensing (see Mandelbaum et al. 2016; Velander et al. 2014; Rodriguez-Puebla et al. 2015; Hudson et al. 2015) and satellite kinematics (see More et al. 2011; Wojtak \& Mamon 2013). We stress that the abundance matching results must be confronted with the data of spheroidal galaxies for stellar masses above, and with the data of disc-dominated galaxies below, a few $10^{10}\, M_\odot$; this is because spheroids and discs mostly contribute to the local stellar mass function in such stellar mass ranges (see Moffett et al. 2016). Provided that, we find a very good agreement, within the uncertainties of the respective datasets.

The dotted green line is instead the outcome when matching the local stellar mass function with the galactic halo mass function. This highlights that the decrease in star-formation efficiency at large stellar masses is somewhat spurious, being related to the fact that the most massive galaxies tend to live at the center of group/cluster halos, which contain a lot of DM. Considering instead only the DM mass belonging to individual galactic halos would imply the efficiency to stay almost constant or increase somewhat at large masses out to $M_\star\sim 10^{12}\, M_\star$.

The resulting values and shape of the star formation efficiency as a function of stellar mass is easily understood in terms of feedback processes. It is apparent that, because of feedbacks, galaxy formation is a very inefficient process: at most $20-30\%$ of the original baryonic content of halos is converted into stars; this occurs for galaxies with final stellar mass around a few $10^{10}\, M_\odot$ (corresponding to halos with mass $M_{\rm H}\approx 10^{12}\, M_\odot$). At small stellar masses, the action of supernova feedback is predominant, while for large stellar masses AGN feedback is likely more relevant; the mass of maximum efficiency corresponds approximately to the transition between supernova and AGN feedbacks (see Shankar et al. 2006; Moster et al. 2013; Aversa et al. 2015).

The green dashed line is the outcome of matching the stellar mass function and the overall halo mass function at $z\approx 2$, and compares well with the efficiencies measured at the same redshift from H$\alpha$ observation by Burkert et al. (2016). The outcome is also similar, within a factor of $2$, to the determination via abundance matching by Moster et al. (2013), Behroozi et al. (2013), and Aversa et al. (2015). The similarity of the efficiency at $z\approx 2$ to the local value is indicative that star formation is mainly an in-situ process (see Lilly et al. 2013; Moster et al. 2013; Aversa et al. 2015; Mancuso et al. 2016a).

We also present as a red solid line the outcome of matching the overall halo mass function with the stellar mass function of passively evolving galaxies, again finding a pleasingly agreement with the local data for spheroidal galaxies. It is extremely interesting to notice that the global efficiency at $z\approx 2$ can be brought on the efficiency of quiescent galaxies at $z\approx 0$ by allowing:  (i) an evolution of the stellar mass by a factor $50\%$ due to late star formation or dry mergers (see Rodriguez-Puebla et al. 2017; also Sect.~\ref{sec|drymerg}); (ii) an halo mass evolution by a factor $4\,(M_{\rm H}/10^{14}\, M_\odot)^{0.12}$ due to late smooth accretion or tidal stripping (see McBride et al. 2009; Fakhouri et al. 2010; Lapi et al. 2013). We stress that such result is again indicative of the in-situ nature of the star formation in spheroid progenitors, and is also extremely relevant for understanding the evolution of the specific angular momentum in galaxies (see Shi et al. 2017).

In Fig.~\ref{fig|mainseq} we show the main sequence of star-forming galaxies at different redshifts, as obtained by matching the SFR function and the stellar mass functions from the continuity equation after Eq.~(\ref{eq|mainseq}). The outcome at $z\approx 2$ is in pleasing agreement with the observational determination from large statistics of mass-selected galaxy samples by Rodighiero et al. (2015).
This further substantiate our assumed star formation histories for individual galaxies, that are illustrated on three representative cases by red dotted lines; their shape is dictated by the slowly increasing SFR $\dot M_\star\propto \tau^{1/2}$ and appreciably rising stellar mass $M_\star\propto \tau^{3/2}$, which imply $\dot M_\star\propto M_\star^{1/3}$. Then the main sequence corresponds to the portions of such tracks where galaxies spend most of their lifetime in
logarithmic bins of $M_\star$.

To highlight the relevance of observational selections different from that based on stellar mass, in Fig.~\ref{fig|mainseq} we also report data points for individual, far-IR selected galaxies by Koprowski et al. (2016), Ma et al. (2015b), Negrello et al. (2014), along with Dye et al. (2015), da Cunha et al. (2015), and Dunlop et al. (2017) mainly at redshifts $z\sim 1-4$.
An appreciable fraction of the individual, far-IR selected galaxies around $z\approx 2$ (highlighted in red) lie above the main sequence, i.e., at SFR values higher than expected on the basis of the average relationship at given $M_\star$. These off-main-sequence objects can be simply interpreted (see Mancuso et al. 2016b) as galaxies caught in an early evolutionary stage, and still accumulating their stellar mass. Thus young star-forming galaxies are found to be preferentially located above the main sequence or, better, to the left of it. As time goes by and stellar mass increases, the galaxy moves toward the average main sequence relationship, around which it will spend most of its lifetime. Afterwards, the SFR is quenched by feedbacks and the galaxy will then evolve passively to become a local early-type; then it will populate a region of the SFR versus stellar mass diagram substantially below the main sequence. These loci of 'red and dead' galaxies are indeed observed locally (see Renzini \& Peng 2015), and start to be pinpointed even at high redshift (see Man et al. 2016).

\section{Summary}\label{sec|summary}

We have developed the continuity equation for the stellar mass content of galaxies, and have exploited it to derive the stellar mass function of active and quiescent galaxies at redshifts $z\sim 0-8$ from the observed SFR functions and disc-like or spheroid-like star-formation histories for individual galaxies. Our approach based on the continuity equation includes a source term due to dry merging gauged on state-of-the-art numerical simulations and consistent with observations. We have then used the abundance matching technique to investigate the star formation efficiency and the main sequence of star-forming and quiescent galaxies. By comparing these outcomes to current observational estimates, we have inferred constraints on the characteristic timescales for star formation and quiescence, on the overall star formation efficiency, and on the amount of stellar mass added by in-situ star formation vs. that contributed by external (dry) merger events.

Our main findings are the following:

\begin{itemize}

\item We have found that the stellar mass function computed from the continuity equation is in excellent agreement with current observational constraints in the extended redshift range $z\sim 0-8$.  At high redshift $z\ga 1$ the mass function is produced by galaxies with spheroid-like star-formation histories, featuring an approximately constant (or slowly increasing) behavior of the SFR as a function of galactic age; the SFR must last for a time $\tau_{\rm sphe}\approx$ fraction of Gyr in strongly star-forming galaxies, while it can proceed over a longer time interval up to a few Gyrs for less massive objects: this reflects the differential action of supernova and AGN feedbacks in systems with different mass. We stressed the relevance of using as input of the continuity equation the SFR function estimated from far-IR/sub-mm/radio, in addition to UV, observations. This is because strongly star-forming galaxies are heavily dust-enshrouded, and as such their intrinsic SFR is considerably underestimated by UV observations, even when corrected for dust extinction according to standard prescriptions based on the UV slope. We have highlighted that the mass growth of spheroids is dominated by in-situ star formation for $z\ga 1$, while at lower redshift dry mergers contribute a mass budget $\la 50\%$ especially in the most massive objects.

\item At low redshift $z\la 1$, we have shown that the stellar mass function of disc-dominated galaxies is well reproduced in our approach when using as input the UV-inferred SFR functions and an exponentially declining SFR history with a long characteristic timescales $\tau_{\rm disc}\approx$ several Gyrs. On the other hand, we have noted that assigning such a disc-like star-formation history to the UV+far-IR/sub-mm/radio SFR functions would considerably overproduce the number of massive discs; this is because obscuration is mild in starforming discs, so the SFR function from dust-corrected UV data must be effectively exploited as input of the continuity equation. The effects of mergers on the stellar mass function of discs are negligible.

\item We have found that the stellar mass function of quiescent galaxies from the continuity equation is in excellent agreement with current observational constraints for $z\la 4$. We thus have demonstrated \emph{quantitatively} via the continuity equation that the dusty, strongly star-forming galaxies recently discovered thanks to wide area far-IR/sub-mm surveys at $z\ga 1$ are indeed the progenitors of the massive quiescent galaxies increasingly detected out to high redshift $z\la 4$ via deep near-IR surveys. We have estimated that the typical time of quiescence (i.e., with absent or negligibly small SFR) for these galaxies is around $\Delta t_{\rm qui}\approx 250-500$ Myr.
    To further test the outcomes of the continuity equation, and better constrain the input SFR functions and the parameters of the star-formation history for individual galaxies, it will be crucial to improve the accuracy in the determination of the stellar mass function at the high-mass end $M_\star\ga 10^{11}\, M_\odot$ out to $z\la 6$ via wide areas near-IR surveys.

\item We have determined the cosmic mass density, finding it in excellent agreement with observational determination out to $z\sim 0-8$, both for active and quiescent galaxies. The continuity equation implies an analogue of the Soltan argument for the stellar component, in such a way that the cosmic stellar mass density is by construction consistent with the cosmic time-integrated star formation history, besides a factor depending on the IMF.

\item We have determined the star-formation efficiency of galaxies as a function of the stellar mass in the local Universe, finding it in good agreement with diverse observations. We have found, in line with previous studies, that the efficiency of star-formation is lower than $f_\star\approx 20-30\%$, with the maximum value being attained around a characteristic stellar mass of a few $10^{10}\, M_\odot$. The behavior as a function of stellar mass can  be ascribed to different form of feedbacks regulating star formation in galaxies, with supernovae and stellar winds dominating for stellar masses below the characteristic one, and AGN feedback dominating above.
    We have also pointed out that the decline of the efficiency for large
    masses is somewhat spurious, being related to the fact that the most
    massive galaxies tend to live at the center of group/cluster halos, which contain a lot of DM; considering instead only the DM mass belonging to individual galactic halos would imply the efficiency to stay almost constant or increase somewhat at large stellar masses. Finally, we have stressed that the similarity of the efficiency at $z\ga 2$ to the local one is indicative of the early, in-situ nature of the star formation process, at least for massive spheroidal galaxies; in fact, we have noted that the global efficiency at $z\approx 2$ can be brought on that observed locally for quiescent galaxies by letting the stellar mass to evolve of a modest factor 50\% due to late star formation or dry mergers and the halo mass to evolve by a factor of a few due to late smooth accretion and/or tidal stripping.

\item We have computed the main sequence of star forming galaxies via abundance matching of the input SFR function and of the stellar mass function self-consistently derived from the continuity equation. We have found a remarkable agreement with the observational determinations at different redshifts, so further constraining our input star-formation histories and timescales. We have highlighted how off-main sequence galaxies (located above the average relation) can be simply interpreted in the light of our star formation histories as
    young objects, caught when their stellar mass is still to be accumulated; they will then progressively move onto the main sequence, where they will spend most of their lifetime as active galaxies, before being quenched.

\end{itemize}

Finally, we conclude by stressing that the added value of the continuity equation, developed here on the stellar component of galaxies, is to provide \emph{quantitative}, yet largely \emph{model-independent} outcomes which must be complied by detailed physical models. In particular, the continuity equation allows a full exploitation of the redshift-dependent SFR functions, stellar mass functions, and galaxy main sequence, in order to determine the average star-formation histories and timescales of individual galaxies. Our analysis highlights that a bimodal star-formation history is required for spheroids and discs: the former must be characterized by a nearly constant SFR over short timescales $\la$ Gyr (increasing somewhat for less star-forming objects), and the latter must feature a SFR exponentially declining over long timescales of several Gyrs. Such outcomes of the continuity equation can provide inspiring hints on ways to improve the (sub-grid) physical recipes implemented in theoretical models and numerical simulations; moreover, they can offer a benchmark for forecasts on future observations at very high redshift with multi-band coverage on medium and wide areas, as it will become routinely achievable with the advent of the JWST.

\begin{acknowledgements}
We are grateful to J. Beacom, K. Glazebrook, M. Massardi, F. Pozzi, and P. Salucci for stimulating discussions. We thank the referee for a constructive report. Work partially supported by PRIN MIUR 2015 `Cosmology and Fundamental Physics: illuminating the Dark Universe with Euclid' and PRIN INAF 2014 `Probing the AGN/galaxy co-evolution through ultra-deep and ultra-high-resolution radio surveys'. AL acknowledges the RADIOFOREGROUNDS grant (COMPET-05-2015, agreement number 687312) of the European Union Horizon 2020 research and innovation programme.
\end{acknowledgements}

\end{document}